\documentclass[singlecolumn,epjc3]{svjour3}
\smartqed  
\RequirePackage{graphicx}
\journalname{Eur. Phys. J. C}

\begin{document}

\title{Possibility of higher dimensional anisotropic compact star}

\author{Piyali Bhar\thanksref{e1,addr1}
\and
        Farook  Rahaman\thanksref{e2,addr2}
       \and
       Saibal Ray\thanksref{e3,addr3}
        \and
        Vikram Chatterjee\thanksref{e4,addr4}.
}

\thankstext{e1}{e-mail: piyalibhar90@gmail.com}
\thankstext{e2}{e-mail: rahaman@iucaa.ernet.in}
\thankstext{e3}{e-mail: saibal@iucaa.ernet.in}
\thankstext{e4}{e-mail: vikphy1979@gmail.com}

\institute{Department of Mathematics, Jadavpur University, Kolkata
700032, West Bengal, India\label{addr1}
              \and
              Department of Mathematics, Jadavpur University, Kolkata
700032, West Bengal, India\label{addr2}
              \and
              Department of Physics, Government College of Engineering \& Ceramic Technology, Kolkata
700010, West Bengal, India\label{addr3}
              \and
               Department of Physics, Central Footwear Training Centre, Kalipur,
Budge Budge, South 24 Parganas 700138, West Bengal,
India\label{addr4} }

\date{Received: date / Accepted: date}

\maketitle

\begin{abstract}
We provide here a new class of interior solutions for anisotropic
stars admitting conformal motion in higher dimensional
noncommutative spacetime. The Einstein field equations are solved
by choosing a particular density distribution function of
Lorentzian type as provided by Nazari and Mehdipour
\cite{Nozari2009,Mehdipour2012} under noncommutative geometry.
Several cases with dimensions $4D$ and higher, e.g. $5D$, $6D$ and
$11D$ have been discussed separately ($D$ stands for dimension of
the spacetime). An overall observation is that the model
parameters, such as density, radial pressure, transverse pressure
and anisotropy all are well behaved and represent a compact star
with mass $2.27$~$M_{\odot}$ and radius $4.17$~km. However,
emphasis has been given on the acceptability of the model from a
physical point of view. As a consequence it is observed that
higher dimensions, i.e. beyond $4D$ spacetime, exhibit several
interesting yet bizarre features which are not at all untenable
for a compact stellar model of strange quark type and thus
dictates a possibility of its extra dimensional existence.
\end{abstract}


\section{Introduction}
To model a compact object it is generally assumed that the
underlying matter distribution is homogeneous i.e.
perfect fluid obeying Tolman-Oppenheimer-Volkoff (TOV) equation.
The nuclear matter of density $\rho \sim 10^{15}$~gm/cc, which is
expected at the core of the compact terrestrial object, becomes
anisotropic in nature as was first argued by Ruderman
\cite{Ruderman1972}. In case of anisotropy the pressure inside
the fluid sphere can specifically be decomposed into two parts:
radial pressure, $p_r$ and the transverse pressure, $p_t$, where $p_t$
is in the perpendicular direction to $p_r$. Their difference
$\Delta=p_t-p_r$ is defined as the anisotropic factor. Now, the
anisotropic force ($\frac{2\Delta}{r}$) will be repulsive in nature
if $\Delta>0$ or equivalently $p_t>p_r$ and attractive if $p_t<p_r$.
So it is reasonable to consider pressure anisotropy to develop
our model under investigation. It has been shown that in case of
anisotropic fluid the existence of repulsive force helps to construct
compact objects \cite{gokhroo1994}.

Anisotropy may occur for different reasons in any stellar distribution.
It could be introduced by the existence of the solid core or
for the presence of type $3A$ superfluid \cite{Kippenhahn1990}.
Different kinds of phase transitions \cite{sokolov1980}, pion condensation
~\cite{sawyer1972} etc. are also reasonable for anisotropy. It may also
occur by the effects of slow rotation in a star. Bowers and Liang
\cite{Bowers1974} showed that anisotropy might have
non-negligible effects on such parameters like equilibrium mass and
surface redshift. Very recently other theoretical advances also indicate
that the pressure inside a compact object is not essentially isotropic in
nature \cite{Varela2010,Rahaman2010a,Rahaman,Rahaman2012a,Kalam2012,Hossein2012,Kalam2013}.

In recent years the extension of General Relativity to higher
dimensions has become a topic of great interest. As a special
mention in this line of thinking we note that `Whether the usual
solar system tests are compatible with the existence of higher
spatial dimensions' has been investigated by Rahaman et
al.~\cite{Rahaman2}. Some other studies in higher dimension are
done by Liu and Overduin \cite{Liu} for the motion of test
particle whereas Rahaman et al.~\cite{Rahaman3} have investigated
higher dimensional gravastars.

One of the most interesting outcomes of string theory is that
the target spacetime coordinates become noncommuting operators on
$D$-brane~\cite{Witten1996,Seiberg1999}. Now the noncommutativity of
a spacetime can be encoded in the commutator
$\left[x^{\mu},x^{\nu}\right]=i \theta^{\mu \nu}$, where
$\theta^{\mu \nu}$ is an anti-symmetric matrix and is of dimension
$(length)^{2}$ which determines the fundamental cell
discretization of spacetime. It is similar to the way the
Planck constant $\hbar$ discretizes phase space~\cite{Smailagic2003}.

In the literature many studies are available on noncommutative
geometry, for example, Nazari and Mehdipour~\cite{Nozari2009} used Lorentzian
distribution to analyze `Parikh-Wilczek Tunneling' from
noncommutative higher dimensional black holes. Besides this
investigation some other noteworthy works are on galactic rotation
curves inspired by a noncommutative-geometry
background~\cite{Rahaman2012b}, stability of a particular class of
thin-shell wormholes in noncommutative geometry~\cite{Peter2012},
higher-dimensional wormholes with noncommutative geometry
\cite{Rahaman2012c}, noncommutative BTZ black hole
\cite{Rahaman2013}, noncommutative wormholes \cite{Rahaman2014a}
and noncommutative wormholes in $f(R)$ gravity with Lorentzian
distribution~\cite{Rahaman2014b}.

It is familiar to search for the natural relationship between
geometry and matter through the Einstein field equations where it
is very convenient to use inheritance symmetry. The well known
inheritance symmetry is the symmetry under conformal Killing
vectors (CKV) i.e.
\begin{equation}
L_{\xi} g_{ik}=\psi g_{ik},
\end{equation}
where $L$ is the Lie derivative of the metric tensor, which
describes the interior gravitational field of a stellar
configuration with respect to the vector field $\xi$, and $\psi$
is the conformal factor. It is supposed that the vector $\xi$
generates the conformal symmetry and the metric $g$ is conformally
mapped onto itself along $\xi$. It is to note that neither $\xi$
nor $\psi$ need to be static even though one considers a static
metric~\cite{Harko1,Harko2}. We also note that (i) if $\psi=0$
then Eq. (1) gives the Killing vector, (ii) if $\psi=$ constant it
gives homothetic vector and (iii) if $\psi=\psi(\textbf{x},t)$
then it yields conformal vectors. Moreover it is to be mentioned
that for $\psi=0$ the underlying spacetime becomes asymptotically
flat which further implies that the Weyl tensor will also vanish.
So CKV provides a deeper insight of the underlying spacetime
geometry.

A large number of works on conformal motion have been done
by several authors. A class of solutions for anisotropic stars admitting
conformal motion have been studied by Rahaman et
al.~\cite{Rahaman7}. In a very recent work Rahaman et
al.~\cite{Rahaman8} have also described conformal motion in higher
dimensional spacetimes. Charged gravastar admitting conformal
motion has been studied by Usmani et al.~\cite{Usmani2011}. Contrary to
this work Bhar \cite{piyali} has studied higher dimensional charged
gravastar admitting conformal motion whereas
relativistic stars admitting conformal motion has been analyzed by
Rahaman et al.~\cite{Rahaman2010b}. Inspired by these earlier works
on conformal motion we are looking forward for a new class of
solutions of anisotropic stars under the framework of
General Relativity inspired by noncommutative geometry
in four and higher dimensional spacetimes.

In the presence of noncommutative geometry there are two different
distributions available in the literature: (a) Gaussian and (b)
Lorentzian \cite{Mehdipour2012}. Though these two mass distributions 
represent similar quantitative aspects, for the present investigation 
we are exploiting a particular Lorentzian-type energy 
density of the static spherically symmetric smeared and
particle-like gravitational source in the multi-dimensional
general form~\cite{Nozari2009,Mehdipour2012}
\begin{equation}
\rho=\frac{M \sqrt\phi}{\pi^{2}(r^{2}+\phi)^{\frac{n+2}{2}}},
\end{equation}
where $M$ is the total smeared mass of the source, $\phi$ is the
noncommutative parameter which bears a minimal width $\sqrt\phi$
and $n$ is positive integer greater than 1. In this approach,
generally known as the {\it noncommutative geometry inspired
model}, via a minimal length caused by averaging noncommutative
coordinate fluctuations cures the curvature singularity in black
holes
\cite{Smailagic2003,Spallucci2006,Banerjee2009,Modesto2010,Nicolini2011}.
It has been argued that it is not required to consider the length
scale of the coordinate noncommutativity to be the same as the
Planck length as the noncommutativity influences appears on a
length scale which can behave as an adjustable parameter
corresponding to that pertinent scale \cite{Mehdipour2012}.

It is interesting to note that Rahaman et al. \cite{Rahaman7}
have found out a new class of interior solutions for anisotropic
compact stars admitting conformal motion under $4D$ framework of GR.
On the other hand, Rahaman et al. \cite{Rahaman8} have studied
different dimensional fluids, higher as well as lower, inspired by
noncommutative geometry with Gaussian distribution of energy density
and have  shown that at $4D$ only one can get a stable configuration
for any spherically symmetric stellar system. However, in the present
work we have extended the work of Rahaman et al. \cite{Rahaman7} to
higher dimensions and that of Rahaman et al. \cite{Rahaman8} to
higher dimensions with energy density in the form of Lorentzian
distribution. In this approach we are able to generalize both
the above mentioned works to show that compacts stars
may exist even in higher dimensions.

In this paper, therefore we use noncommutative geometry inspired model
to combine the microscopic structure of spacetime with the relativistic
description of gravity. The plan of the present investigation is as follows:
in Section 2 we formulate the Einstein field equations for the interior
spacetime of the anisotropic star. In Section 3 we solve the
Einstein field Equations by using the density function of
Lorentzian distribution type in higher dimensional spacetime as
given by Nozari and Mehdipour~\cite{Nozari2009}. We consider
the cases $n=~2,~3,~4$ and $9$ i.e. $4D,~5D,~6D$ and $11D$
spacetimes in Section 4 to examine expressions for physical
parameters whereas the matching conditions are provided in Section 5.
Various physical properties are explored in Section 6 with
interesting features of the model and present them with
graphical plots for comparative studies among the results of
different dimensional spacetimes. Finally we complete the paper
with some concluding remarks in Section 7.

\section{The interior spacetime and the Einstein field equations}
To describe the static spherically symmetric spacetime (in
geometrical units $G=1=c$ here and onwards) in higher dimension the
line element can be given in the standard form
\begin{equation}
ds^{2}=-e^{\nu(r)}dt^{2}+e^{\lambda(r)}dr^{2}+r^{2}d\Omega_n^{2},
\end{equation}
where
\begin{equation}
d\Omega_n^{2}=d\theta_1^{2}+\sin^{2}\theta_1d\theta_2^{2}+
\sin^{2}\theta_1\sin^{2}\theta_2d\theta_3^{2}+...+\prod_{j=1}^{n-1}\sin^{2}\theta_jd\theta_n^{2},
\end{equation}
where $\lambda$, $\nu$ are functions of the radial coordinate $r$.
Here we have used the notation $D=n+2$,~$D$ is the dimension of
the spacetime.

The energy momentum tensor for the matter distribution can be
taken in its usual form \cite{Lobo}
\begin{equation}
T_{\nu}^{\mu}=(\rho+p_r)u^{\mu}u_{\nu}-p_rg^{\mu}_{\nu}+(p_t-p_r)\eta^{\mu}\eta_{\nu},
\end{equation}
with $u^{\mu}u_{\mu}=-\eta^{\mu}\eta_{\mu}=1$ and
$u^{\mu}\eta_{\nu}= 0$. Here the vector $u^{\mu}$ is the fluid
$(n+2)$-velocity and $\eta^{\mu}$ is the unit space-like vector
which is orthogonal to $ u^{\mu}$, where $\rho$ is the matter
density, $p_r$ is the radial pressure in the direction of
$\eta^{\mu}$ and $p_t$ is the transverse pressure in the
orthogonal direction to $p_r$. Since the pressure is anisotropic
in nature so for our model $p_r \neq p_t$. Here $p_t-p_r = \Delta$
is the measure of anisotropy, as defined earlier.

Now for higher ( $n  \geq 2 )$ dimensional spacetime the Einstein
equations can be written as \cite{Rahaman2012b}
\begin{equation}
e^{-\lambda}\left[\frac{n\lambda'}{2r}-\frac{n(n-1)}{2r^{2}}\right]+\frac{n(n-1)}{2r^{2}}=8\pi
\rho=8\pi T_0^{0},
\end{equation}

\begin{equation}
e^{-\lambda}\left[\frac{n(n-1)}{2r^{2}}+\frac{n\nu'}{2r}\right]-\frac{n(n-1)}{2r^{2}}=8\pi
p_r=-8\pi T_1^{1},
\end{equation}

\[\frac{1}{2}e^{-\lambda}\left[\frac{1}{2}(\nu')^{2}+\nu''-\frac{1}{2}\lambda'\nu'+\frac{(n-1)}{r}(\nu'-\lambda')
+\frac{(n-1)(n-2)}{r^{2}}\right]\]
\begin{equation}
-\frac{(n-1)(n-2)}{2r^{2}}=8\pi p_t=-8\pi T_2^{2}=-8\pi T_3^{3},
\end{equation}
where $\prime$ denotes differentiation with respect to the radial
coordinate $r$ i.e. $\prime \equiv \frac{d}{dr}$.

\section{The solution under conformal Killing vector}
Mathematically, conformal motions or conformal Killing vectors
(CKV) are motions along which the metric tensor of a spacetime
remains invariant up to a scale factor. A conformal vector field
can be defined as a global smooth vector field $x$ on a manifold,
$\ss$, such that for the metric $g_{ab}$ in any coordinate system
on $\ss$ $x_{a;b} = \psi g_{ab} + F_{ab}$, where $\psi : \ss
\rightarrow real ~number$ is the smooth conformal function of
$x$, $F_{ab}$ is the conformal bivector of $x$. This is
equivalent to $L_x g_{ik} = \psi g_{ik}$, (as considered in Eq.
(1) in the usual form) where $L_x$ signifies the Lie derivatives
along $x_a$.

To search the natural relation between geometry and matter through
the Einstein equations, it is useful to use inheritance symmetry.
The well known inheritance symmetry is the symmetry under
conformal Killing vectors (CKV). These provide a deeper insight into
the spacetime geometry. The CKV facilitate generation
of exact solutions to the Einstein's field equations. The study of
conformal motions in spacetime is physically very important
because it can lead to the discovery of conservation laws and devise
spacetime classification schemes. Einstein's field equations being
highly non linear partial differential equations, one can reduce
the partial differential equations to ordinary differential equations
by use of CKV. It is still a challenging problem to the theoretical
physicists to know the exact nature and characteristics of compact stars
and elementary particle like electron.

Let us therefore assume that our static spherically symmetry spacetime
admits an one parameter group of conformal motion. The conformal Killing vector,
as given in Eq. (1), can be written in a more convenient form:
\begin{equation}
L_\xi g_{ik}=\xi_{i;k}+\xi_{k;i}=\psi g_{ik},
\end{equation}
where both $i$ and $k$ take the values $  ~1,~2...,~n+2$. Here
$\psi$ is an arbitrary function of the radial coordinate $r$ and
$\xi$ is the orbit of the group. The metric $g_{ij}$ is
conformally mapped onto itself along $\xi_i$.

Let us further assume that the orbit of the group to be orthogonal
to the velocity vector field of the fluid,
\begin{equation}
\xi^{\mu}u_{\mu}=0.
\end{equation}

As a consequence of the spherically symmetry from Eq. (10) we have
\[\xi^{1}=\xi^{3}=...=\xi^{n+1}=0.\] Now the conformal Killing equation
for the line element (3) gives the following equations:
\begin{equation}
\xi^{2}\nu'=\psi,
\end{equation}

\begin{equation}
\xi^{n+2}=C_1,
\end{equation}

\begin{equation}
\xi^{2}=\frac{\psi r}{2},
\end{equation}

\begin{equation}
\xi^{2}\lambda'+2{\xi^{2}}^\prime =\psi,
\end{equation}
where $2$ stands for the spatial coordinates $r$,`$\prime$' and
`$,$' denotes the partial derivative with respect to $r$ and $C_1$
is a constant.

The above set of equations consequently gives
\begin{equation}
e^{\nu}=C_2^{2}r^{2},
\end{equation}

\begin{equation}
e^{\lambda}=\left(\frac{C_3}{\psi}\right)^{2},
\end{equation}

\begin{equation}
\xi^{i}=C_1\delta_{n+2}^{i}+\left( \frac{\psi
r}{2}\right)\delta_2^{i},
\end{equation}
where $\delta$ stands for the `Kronecker delta'
and $C_2$,~$C_3$ are constants of integrations.

Using Eqs. (15)-(17) in the Einstein field Eqs. (6)-(8), we get
\begin{equation}
\frac{n(n-1)}{2r^{2}}\left(1-\frac{\psi^{2}}{C_3^{2}}\right)-\frac{n\psi
\psi'}{rC_3^{2}}=8\pi \rho,
\end{equation}

\begin{equation}
\frac{n}{2r^{2}}\left[(n+1)\frac{\psi^{2}}{C_3^{2}}-(n-1)\right]=8\pi
p_r,
\end{equation}

\begin{equation}
\frac{n\psi\psi'}{rC_3^{2}}+n(n-1)\frac{\psi^{2}}{2r^{2}C_3^{2}}-\frac{(n-1)(n-2)}{2r^{2}}=8\pi
p_t.
\end{equation}

We thus have three independent Eqs. (18)-(20)
with four unknowns $\rho$,~$p_r$,~$p_t$,~$\psi$. So we
are free to choose any physically reasonable {\it ansatz} for any
one of these four unknowns. Hence we choose density profile $\rho$ in
the form given in Eq. (2) in connection to higher dimensional
static and spherically symmetric Lorentzian distribution of
smeared matter as provided by Nozari and Mehdipour~\cite{Nozari2009}.
This density profile will be employed as a key tool in our present
study.

Therefore, substituting Eq. (2) into (18) and solving, we
obtain
\begin{equation}
\psi^{2}=C_3^{2}-\frac{16MC_3^{2}\sqrt{\phi}}{n\pi}\frac{1}{r^{n-1}}\int\frac{r^{n}}{(r^{2}+\phi)^{\frac{n+2}{2}}}dr+\frac{A}{r^{n-1}},
\end{equation}
where $A$ is a constant of integration which is determined by
invoking suitable boundary conditions.

Now the equation (21) gives the expression of the conformal factor $\psi$.
Assigning $n=~2,~3,~4$ and $9$ i.e. $4D,~5D,~6D$ and $11D$ spacetimes
respectively if we perform the above integral then the conformal factor
$\psi$ can be obtained for different dimension that is necessary to
find out the other physical parameter namely $p_r$ and $p_t$ for these dimension.
Here $A,~-\infty <A <\infty$, is a constant of integration that can be later on found
out from the boundary condition $p_r(R)=0$,~$R$ being the radius of the star.

\section{Exact solutions of the models in different dimensions}
The above set of equations are associated with dimensional
parameter $n$ and hence to get a clear picture of the physical
system under different spacetimes we are interested for studying
several cases starting from standard $4D$ to higher $5D$, $6D$
and $11D$ spacetimes as shown below.

\subsection{Four dimensional spacetime ($n=2$)}
The conformal parameter $\psi(r)$ and the metric potential
$e^{\lambda}$ are given as,
\begin{equation}
\psi=\sqrt{C_3^{2}-\frac{4MC_3^{2}\sqrt{\phi}}{\pi
r}\left[\frac{1}{\sqrt{\phi}}\arctan\left(\frac{r}{\sqrt{\phi}}\right)
-\frac{r}{r^{2}+\phi} \right]+\frac{A}{r}},
\end{equation}

\begin{equation}
e^{-\lambda}=1+\frac{A}{C_3^{2}r}-\frac{4M\sqrt{\phi}}{\pi
r}\left[\frac{1}{\sqrt{\phi}}\arctan\left(\frac{r}{\sqrt{\phi}}\right)
-\frac{r}{r^{2}+\phi}\right].
\end{equation}

The radial and transverse pressures are obtained as
\begin{equation}
p_r=\frac{1}{8\pi
r^{2}}\left[2+\frac{3A}{C_3^{2}r}-\frac{12M\sqrt{\phi}}{\pi
r}\left\{\frac{1}{\sqrt{\phi}}\arctan\left(\frac{r}{\sqrt{\phi}}\right)
-\frac{r}{r^{2}+\phi} \right\} \right],
\end{equation}

\begin{equation}
p_t=\frac{1}{8\pi}\left[\frac{1}{r^{2}}-\frac{8M\sqrt{\phi}}{\pi(r^{2}+\phi)^{2}}\right].
\end{equation}

To find the above constant of integration we impose the boundary
condition $p_r(r=R)=0$, where $R$ is the radius of the fluid
sphere as mentioned earlier, which gives
\begin{equation}
A=\frac{4MC_3^{2}\sqrt{\phi}}{\pi}\left\{
\frac{1}{\sqrt{\phi}}\arctan\left(\frac{R}{\sqrt{\phi}}\right)
-\frac{R}{R^{2}+\phi}\right\}-\frac{2}{3}C_3^{2}R.
\end{equation}

\subsection{Five dimensional spacetime ($n=3$)}
In this case the solution set can be obtained as follows:
\begin{equation}
\psi=\sqrt{C_3^{2}+\frac{16MC_3^{3}\sqrt{\phi}}{9\pi
r^{2}}\frac{3r^{2}+2\phi}{(r^{2}+\phi)^{\frac{3}{2}}}+\frac{A}{r^{2}}},
\end{equation}

\begin{equation}
e^{-\lambda}=1+\frac{16M\sqrt{\phi}}{9\pi
r^{2}}\frac{3r^{2}+2\phi}{(r^{2}+\phi)^{\frac{3}{2}}}+\frac{A}{C_3^{2}r^{2}},
\end{equation}

\begin{equation}
p_r=\frac{3}{8\pi r^{2}}\left[1+\frac{32M\sqrt{\phi}}{9\pi
r^{2}}\frac{3r^{2}+2\phi}{(r^{2}+\phi)^{\frac{3}{2}}}+\frac{2A}{C_3^{2}r^{2}}\right],
\end{equation}

\begin{equation}
p_t=\frac{1}{8\pi}\left[\frac{2}{r^{2}}-\frac{8M\sqrt{\phi}}{\pi(r^{2}+\phi)^{\frac{5}{2}}}\right],
\end{equation}
with
\begin{equation}
A=-\frac{C_3^{2}R^{2}}{2}\left[1+\frac{32M\sqrt{\phi}}{9\pi
R^{2}}\frac{3R^{2}+2\phi}{(R^{2}+\phi)^{\frac{3}{2}}}\right].
\end{equation}

\subsection{Six dimensional spacetime ($n=4$)}
Here the solutions are as follows:
\begin{equation}
\psi=\sqrt{C_3^{2}-\frac{MC_3^{2}\sqrt{\phi}}{2\pi
r^{3}}\left[\frac{3}{\sqrt{\phi}}\tan^{-1}\left(\frac{r}{\sqrt{\phi}}\right)
-\frac{5r^{3}+3r\phi}{(r^{2}+\phi)^{2}}
\right]+\frac{A}{r^{3}}},
\end{equation}

\begin{equation}
e^{-\lambda}=1-\frac{M \sqrt{\phi}}{2\pi
r^{3}}\left[\frac{3}{\sqrt{\phi}}\tan^{-1}\left(\frac{r}{\sqrt{\phi}}\right)
-\frac{5r^{3}+3r\phi}{(r^{2}+\phi)^{2}}
\right]+\frac{A}{C_3^{2}r^{3}},
\end{equation}

\begin{equation}
p_r=\frac{1}{4 \pi r^{2}}\left[2-\frac{5M\sqrt{\phi}}{2\pi
r^{3}}\left\{\frac{3}{\sqrt{\phi}}\tan^{-1}\left(\frac{r}{\sqrt{\phi}}\right)
-\frac{5r^{3}+3r\phi}{(r^{2}+\phi)^{2}}
\right\}+\frac{5A}{r^{3}C_3^{2}}\right],
\end{equation}

\begin{equation}
p_t=\frac{1}{8\pi}\left[\frac{3}{r^{2}}-\frac{8M\sqrt{\phi}}{\pi(r^{2}+\phi)^{3}}\right],
\end{equation}

and

\begin{equation}
A=\frac{MC_3^{2}\sqrt{\phi}}{2\pi}\left\{\frac{3}{\sqrt{\phi}}\tan^{-1}\left(\frac{R}{\sqrt{\phi}}\right)
-\frac{5R^{3}+3r\phi}{(R^{2}+\phi)^{2}}
\right\}-\frac{2C_3^{2}R^{3}}{5}.
\end{equation}

\subsection{Eleven dimensional spacetime ($n=9$)}
For this arbitrarily chosen higher dimension the solutions can be
obtained as
\begin{equation}
\psi=\sqrt{C_3^{2}+C_3^{2}\frac{16M\sqrt{\phi}}{2835\pi
r^{8}}\frac{315r^{8}+840r^{6}\phi+1008r^{4}\phi^{2}+576r^{2}\phi^{3}+128\phi^{4}}{(r^{2}
+\phi)^{\frac{9}{2}}}+\frac{A}{r^{8}}},
\end{equation}

\begin{equation}
e^{-\lambda}=1+\frac{16M\sqrt{\phi}}{2835\pi
r^{8}}\frac{315r^{8}+840r^{6}\phi+1008r^{4}\phi^{2}+576r^{2}\phi^{3}+128\phi^{4}}{(r^{2}
+\phi)^{\frac{9}{2}}}+\frac{A}{C_3^{2}r^{8}},
\end{equation}

\begin{equation}
p_r=\frac{9}{8\pi r^{2}}\left[1+\frac{16M\sqrt{\phi}}{567\pi
r^{8}}\frac{315r^{8}+840r^{6}\phi+1008r^{4}\phi^{2}+576r^{2}\phi^{3}+128\phi^{4}}{(r^{2}
+\phi)^{\frac{9}{2}}}+\frac{5A}{C_3^{2}r^{8}}\right],
\end{equation}

\begin{equation}
p_t=\frac{1}{\pi}\left[\frac{1}{r^{2}}-\frac{M\sqrt{\phi}}{\pi(r^{2}+\phi)^{\frac{11}{2}}}\right],
\end{equation}

and

\begin{equation}
A=-\frac{C_3^{2}}{5}\left[R^{8}+\frac{16M\sqrt{\phi}}{567\pi}\frac{315R^{8}+840R^{6}\phi+1008R^{4}\phi^{2}
+576R^{2}\phi^{3}+128\phi^{4}}{(R^{2}
+\phi)^{\frac{9}{2}}}\right].
\end{equation}

\begin{figure}[h]
\centering
\includegraphics[width=0.4\textwidth]{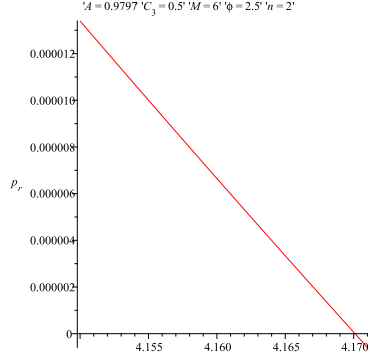}
\caption{The graphical plot for radial pressure vs radius which
has a definite cut-off at $4.17$~km}
\end{figure}

Let us now turn our attention to the physical analysis of the stellar
model under consideration, i.e. whether it is a normal star or
something else. To do so, primarily we try to
figure out the radius of the stellar configuration. It is to note
that in Eqs. (29), (34) and (39) the radius of the star $R$ has
been mentioned under the boundary condition, $p_r(R) =0$, i.e. we
get analytical results in the respective cases. So it seems that
we can proceed on without further plot descriptions to get $p_r(R)
=0$ for all dimensions. However, for $4D$ case it reveals that the
radius of the star is very small with a numerical value of
$4.17$~km (Fig. 1). This obviously then indicates that the star is
nothing but a compact object (see Table 1 of all the Refs.
 \cite{Rahaman,Kalam2012,Hossein2012,Kalam2013}
for comparison with the radius of some of the real compact stars).

\section{Matching conditions}

Now, we match our interior solutions with the  exterior vacuum
solutions. The generalization of Schwarzschild solution, which as
obtained by Tangherlini \cite{Tangherlini1963} reads as

\begin{equation}
ds ^2 = - \left(1 - \frac{\mu_{n}}{ r^{n-1}}\right)  dt^2 +
\left(1 - \frac{\mu_{n}}{ r^{n-1}}\right)^{-1} dr^2 + d
\Omega_{n} ^2.\label{eq16}
\end{equation}

Here, \[ {\Omega}_{ n }=\frac{2 \pi^{\frac{n+1}{2}}}{\Gamma\left(
\frac{n+1}{2}\right)},\] the area of a unit $n$-sphere and
\[\mu_{n}=16\pi GM/n c^2 {\Omega}_{n},\]
 is the constant of integration with $M$, the  mass of the black hole with $n=2,~3,~4,~9$.

\subsection{Four dimensional spacetime ($n=2$)}
For $4D$ case, our interior solution should match to the
exterior Schwarzschild spacetime at the boundary $r=a_4$ given by
\begin{equation}
ds^{2}=-\left(1-\frac{\mu_2}{r}\right)dt^{2}+\left(1-\frac{\mu_2}
{r}\right)^{-1}dr^{2}+r^{2}d\Omega_{2}.
\end{equation}

Now using the matching conditions at the boundary $r=a_4$, we have
\begin{equation}
1-\frac{\mu_2}{a_4}=C_2^{2}a_4^{2},
\end{equation}
and
\begin{equation}
1-\frac{\mu_2}{a_4}=1+\frac{A}{C_3^{2}a_4}-\frac{4M\sqrt{\phi}}{\pi
a_4}\left[\frac{1}{\sqrt{\phi}}\arctan\left(\frac{a_4}{\sqrt{\phi}}\right)
-\frac{a_4}{a_4^{2}+\phi}\right].
\end{equation}

Solving the above two equations, we obtain
\begin{equation}
C_2^{2}=\frac{1}{a_4^{2}}\left(1-\frac{\mu_2}{a_4}\right),
\end{equation}

\begin{equation}
\frac{A}{C_3^{2}}=\frac{4M\sqrt{\phi}}{\pi
}\left[\frac{1}{\sqrt{\phi}}\arctan\left(\frac{a_4}{\sqrt{\phi}}\right)-\frac{a_4}{a_4^{2}+\phi}\right]-\mu_2.
\end{equation}

\subsection{Five dimensional spacetime ($n=3$)}
For $5D$ case, our interior solution should match to the exterior $5D$
Schwarzschild spacetime at the boundary $r=a_5$, given by
\begin{equation}
ds^{2}=-\left(1-\frac{\mu_3}{r^{2}}\right)dt^{2}+\left(1-\frac{\mu_3}{r^{2}}\right)^{-1}dr^{2}
+r^{2}d\Omega_{3}.
\end{equation}

Now using the matching conditions at the boundary $r=a_5$, we have
\begin{equation}
1-\frac{\mu_3}{a_5^{2}}=C_2^{2}a_5^{2},
\end{equation}
and
\begin{equation}
1-\frac{\mu_3}{a_5^{2}}=1+\frac{A}{C_3^{2}a_5^{2}}+\frac{16M\sqrt{\phi}}{9\pi
a_5^{2}}\frac{3a_5^{2}+2\phi}{(a_5^{2}+\phi)^{\frac{3}{2}}}.
\end{equation}

Solving the above two equations, we obtain
\begin{equation}
C_3^{2}=\frac{1}{a_5^{2}}\left(1-\frac{\mu_3}{a_5^{2}}\right),
\end{equation}
and
\begin{equation}
\frac{A}{C_3^{2}}=-\mu_3-\frac{16M\sqrt{\phi}}{9\pi}\frac{3a_5^{2}+2\phi}{(a_5^{2}
+\phi)^{\frac{3}{2}}}.
\end{equation}

\subsection{Six dimensional spacetime ($n=4$)}
For $6D$ case, our interior solution should match to the exterior $6D$
Schwarzschild spacetime at the boundary $r=a_6$, given by
\begin{equation}
ds^{2}=-\left(1-\frac{\mu_4}{r^{3}}\right)dt^{2}+\left(1-\frac{\mu_4}{r^{3}}\right)^{-1}dr^
{2}+r^{2}d\Omega_{4}.
\end{equation}

Now using the matching conditions at the boundary $r=a_6$, we have
\begin{equation}
1-\frac{2\mu_4}{a_6^{3}}=C_2^{2}a_6^{2},
\end{equation}
and
\begin{equation}
1-\frac{\mu_4}{a_6^{3}}=1-\frac{M \sqrt{\phi}}{2\pi
a_6^{3}}\left[\frac{3}{\sqrt{\phi}}\tan^{-1}\left(\frac{a_6}{\sqrt{\phi}}\right)
-\frac{5a_6^{3}+3a_6\phi}{(a_6^{2}+\phi)^{2}}
\right]+\frac{A}{C_3^{2}a_6^{3}}.
\end{equation}

Solving the above two equations, we obtain
\begin{equation}
C_2^{2}=\frac{1}{a_6^{2}}\left(1-\frac{\mu_4}{a_6^{3}}\right),
\end{equation}
and
\begin{equation}
\frac{A}{C_3^{2}}=\frac{M
\sqrt{\phi}}{2\pi}\left[\frac{3}{\sqrt{\phi}}\tan^{-1}\left(\frac{a_6}{\sqrt{\phi}}\right)
-\frac{5a_6^{3}+3a_6\phi}{(a_6^{2}+\phi)^{2}} \right]-\mu_4.
\end{equation}

\subsection{Eleven dimensional spacetime ($n=9$)}
For $11D$ case, our interior solution should match to the exterior
11D Schwarzschild spacetime at the boundary $r=a_{11}$, given by
\begin{equation}
ds^{2}=-\left(1-\frac{\mu_{9}}{r^{8}}\right)dt^{2}+\left(1-\frac{\mu_{9}}{r^{8}}\right)^{-1}
dr^{2}+r^{2}d\Omega_{9}.
\end{equation}

Now using the matching conditions at the boundary $r=a_{11}$, we have
\begin{equation}
1-\frac{\mu_{9}}{a_{11}^{8}}=C_2^{2}a_{11}^{2},
\end{equation}
and
\begin{equation}
1-\frac{\mu_9}{a_{11}^{8}}=1+\frac{16M\sqrt{\phi}}{2835\pi
a_{11}^{8}}\frac{315a_{11}^{8}+840a_{11}^{6}\phi+1008a_{11}^{4}\phi^{2}
+576a_{11}^{2}\phi^{3}+128\phi^{4}}{(a_{11}^{2}
+\phi)^{\frac{9}{2}}}+\frac{A}{C_3^{2}a_{11}^{8}}.
\end{equation}

Solving the above two equations, we obtain
\begin{equation}
C_2^{2}=\frac{1}{a_{11}^{2}}\left(1-\frac{\mu_{9}}{a_{11}^{8}}\right),
\end{equation}
and
\begin{equation}
\frac{A}{C_3^{2}}=\frac{M
\sqrt{\phi}}{2\pi}\left[\frac{3}{\sqrt{\phi}}\tan^{-1}\left(\frac{a_{11}}{\sqrt{\phi}}\right)
-\frac{5a_{11}^{3}+3a_{11}\phi}{(a_{11}^{2}+\phi)^{2}}
\right]-\mu_{9},
\end{equation}
where $a_j$  (j=4,~5,~6~and~11) are the radii of the fluid spheres
in different dimensions.

\section{A comparative study of the physical features of the model}
Let us now carry out a comparative study of the physical features
based on the solutions set obtained in the previous Section 4.
This can be done in different ways. However, in the present
investigation the best method we may adopt for comparative study,
firstly, in connection to stability of the models for different
dimensions which may be considered as most crucial one and
secondly, for other physical parameters viz., density, pressure,
pressure anisotropy, pressure gradient, conformal parameter and
metric potential.

\subsection{Stability of the stellar configuration}
The Generalized Tolman-Oppenheimer-Volkoff (TOV) equation can be
written in the form
\begin{equation}
-\frac{M_G(r)(\rho+p_r)}{r^2}e^{\frac{\nu-\mu}{2}}-\frac{dp_r}{dr}+\frac{2}{r}(p_t-p_r)=0,
\end{equation}
where $M_G(r) $ is the gravitational mass within the sphere of radius $r$
and is given by
\begin{equation}
M_G(r)=\frac{1}{2}r\nu'e^{\frac{\mu-\nu}{2}}.
\end{equation}

Substituting (64) into (63), we obtain
\begin{equation}
-\frac{\nu'}{2}(\rho+p_r)-\frac{dp_r}{dr}+\frac{2}{r}(p_t-p_r)=0.
\end{equation}

The above TOV equation describe the equilibrium of the stellar
configuration under gravitational force $F_g$, hydrostatic force
$F_h$ and anisotropic stress $F_a$ so that we can write it in the
following form:
\begin{equation}
F_g+F_h+F_a=0,
\end{equation}
where
\[F_g=-\frac{\nu'}{2}(\rho+p_r),\]
\[F_h=-\frac{dp_r}{dr},\]
\[F_a=\frac{2}{r}(p_t-p_r).\]
\begin{equation}
\end{equation}

\begin{figure}[h]
\centering
\includegraphics[width=0.4\textwidth]{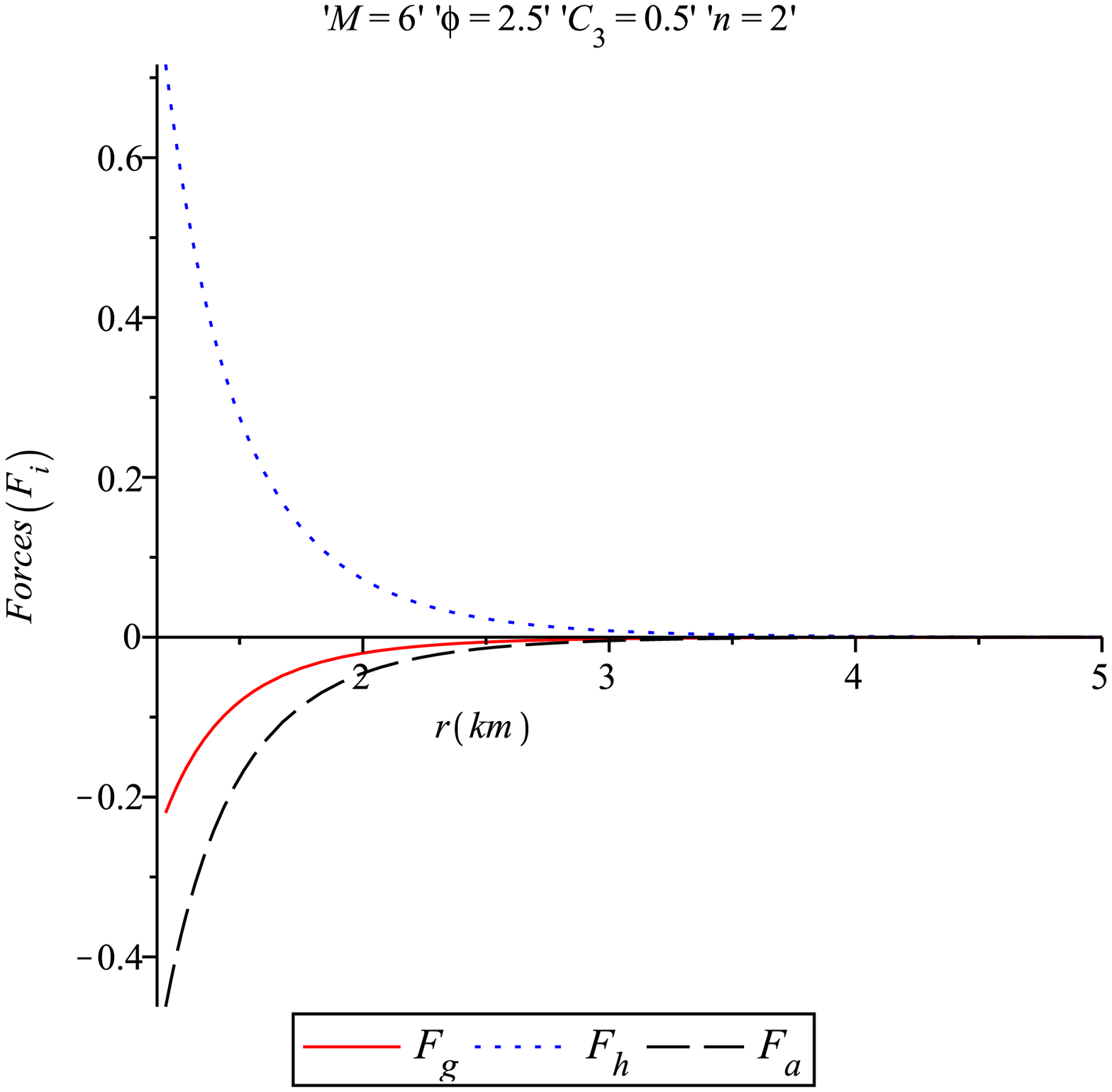}
\includegraphics[width=0.4\textwidth]{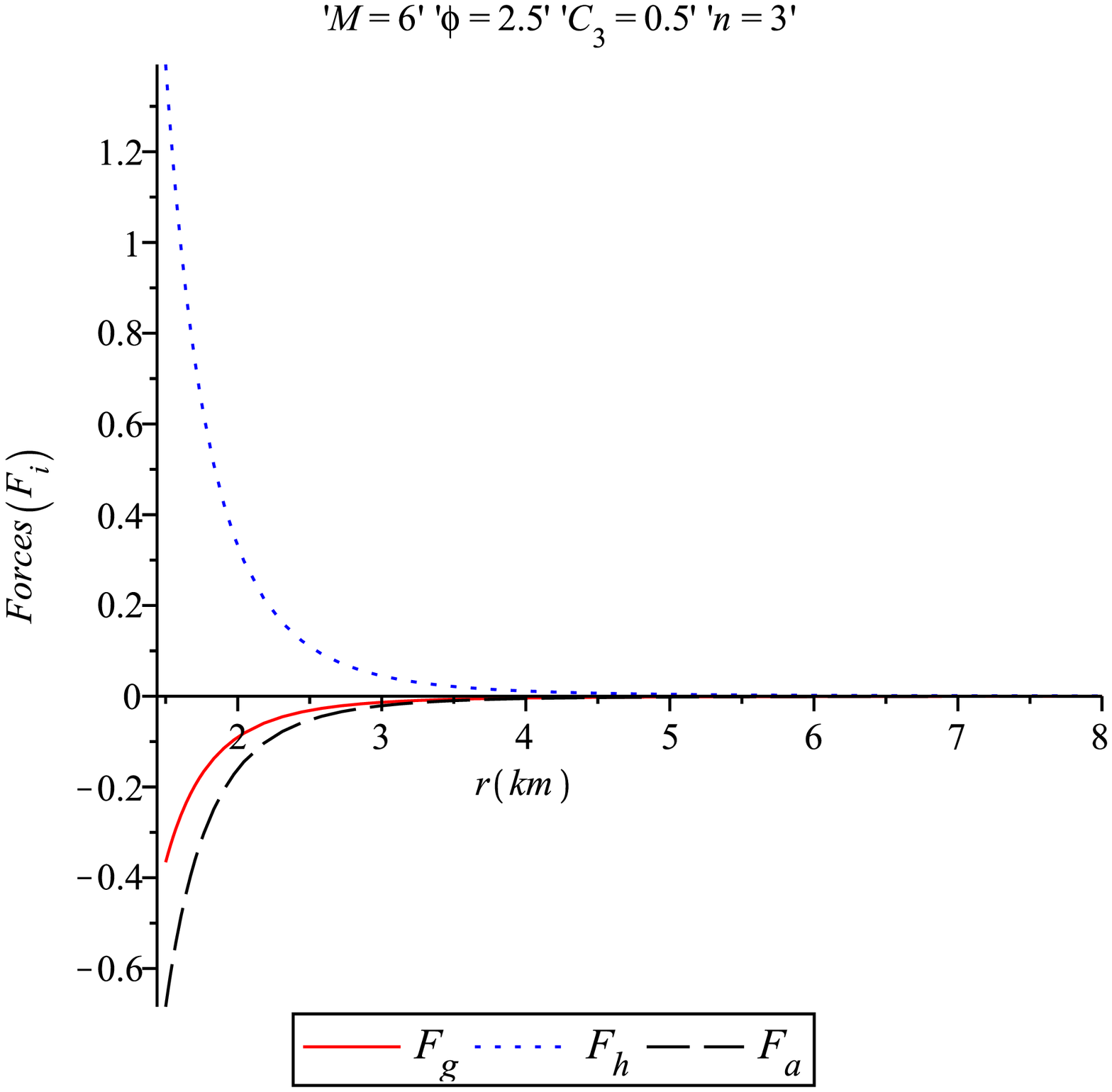}
\includegraphics[width=0.4\textwidth]{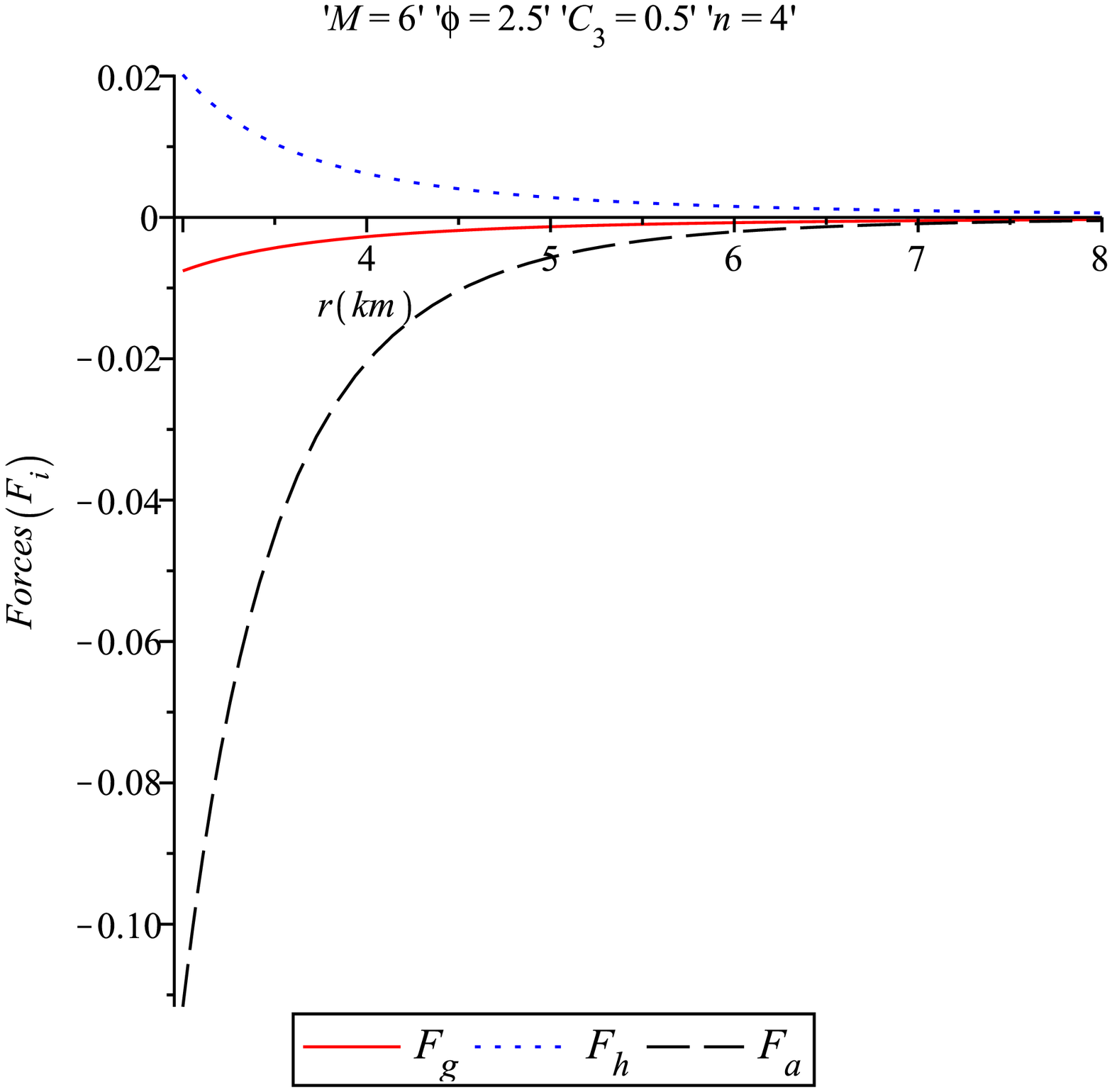}
\includegraphics[width=0.4\textwidth]{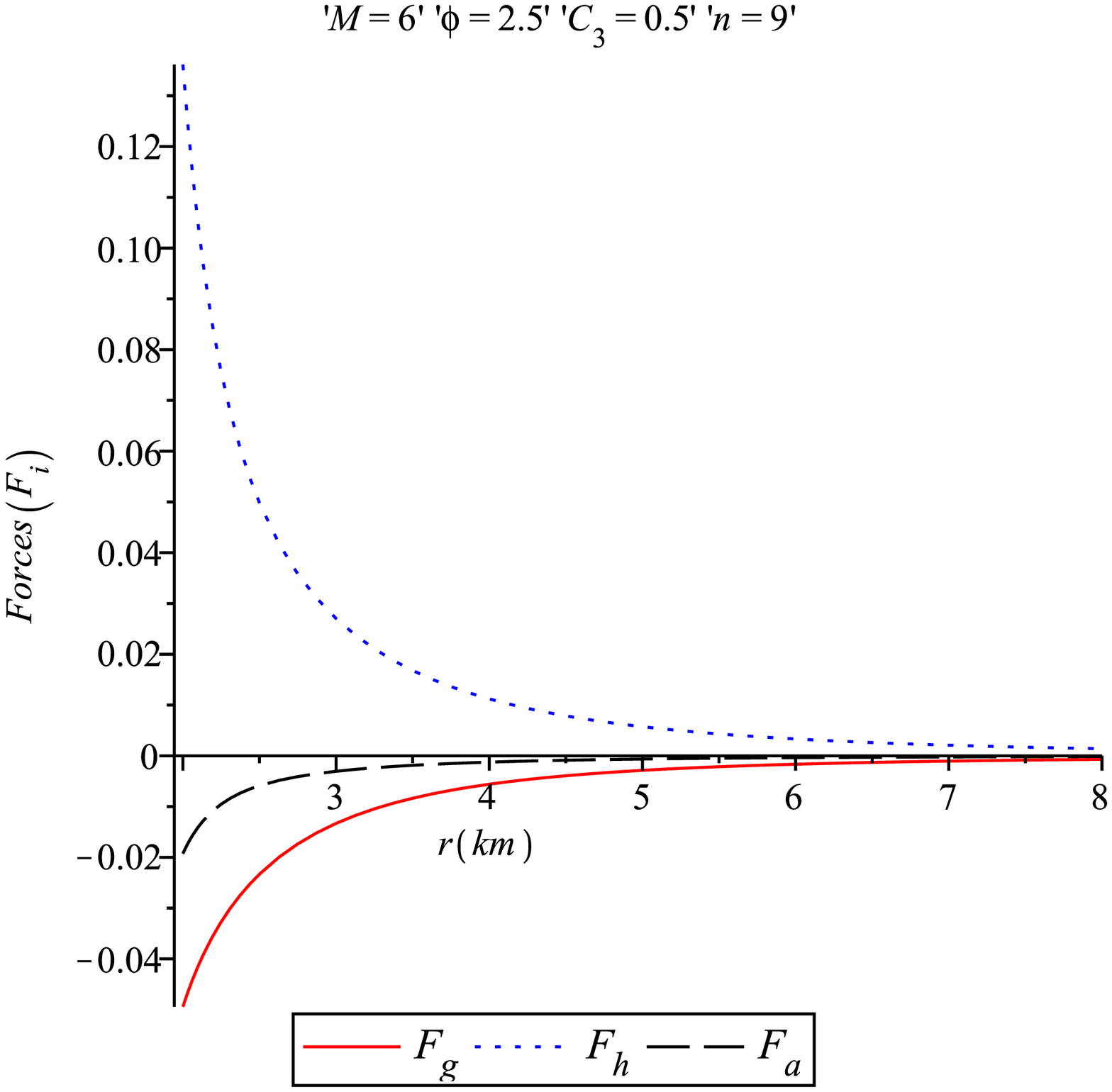}
\caption{The three different forces namely gravitational forces
$(F_g)$, hydrostatic forces $(F_h)$ and anisotropic forces $(F_a)$
are plotted against $r$~(km) for $4D$ spacetime (top left), $5D$
spacetime (top right), $6D$ spacetime (bottom left) and $11D$
spacetime (bottom right). From the figure balancing status of the
forces can be clearly observed for $4D$,~$5D$,~$6D$,~$11D$
spacetimes}
\end{figure}

We have shown the plots of TOV equations for $4D$, $5D$, $6D$ and
$11D$ spacetime in Fig. 2. From the plots it is overall clear that
the system is in static equilibrium under three different forces, viz.
gravitational, hydrostatic and anisotropic, for example, in the $4D$ case
to attain equilibrium, the hydrostatic force is counter
balanced jointly by gravitational and anisotropic forces. In $5D$
also the situation is exactly same, the only difference being in
the radial distances. In $4D$ it is closer to 5 whereas in
$5D$ it is closer to 8. This distance factor can
also be observed in the higher dimensional spacetimes
though the balancing features between the three forces are clearly
different in the respective cases.

\subsection{Energy conditions}
Now we check whether all the energy conditions are satisfied or
not. For  this purpose, we shall consider the following inequalities:
\[
(i)~NEC: \rho+p_r\geq 0,~\rho+p_t\geq 0,
\]
\[
(ii)~WEC: \rho+p_r\geq 0,~\rho\geq 0,~\rho+p_t\geq 0,
\]
\[
(iii)~SEC: \rho+p_r\geq 0,~\rho+p_r+2p_t\geq 0.
\]

Fig. 3 indicates that in our model all the energy conditions are
satisfied through out the interior region.

\begin{figure}[h]
\centering
\includegraphics[width=0.4\textwidth]{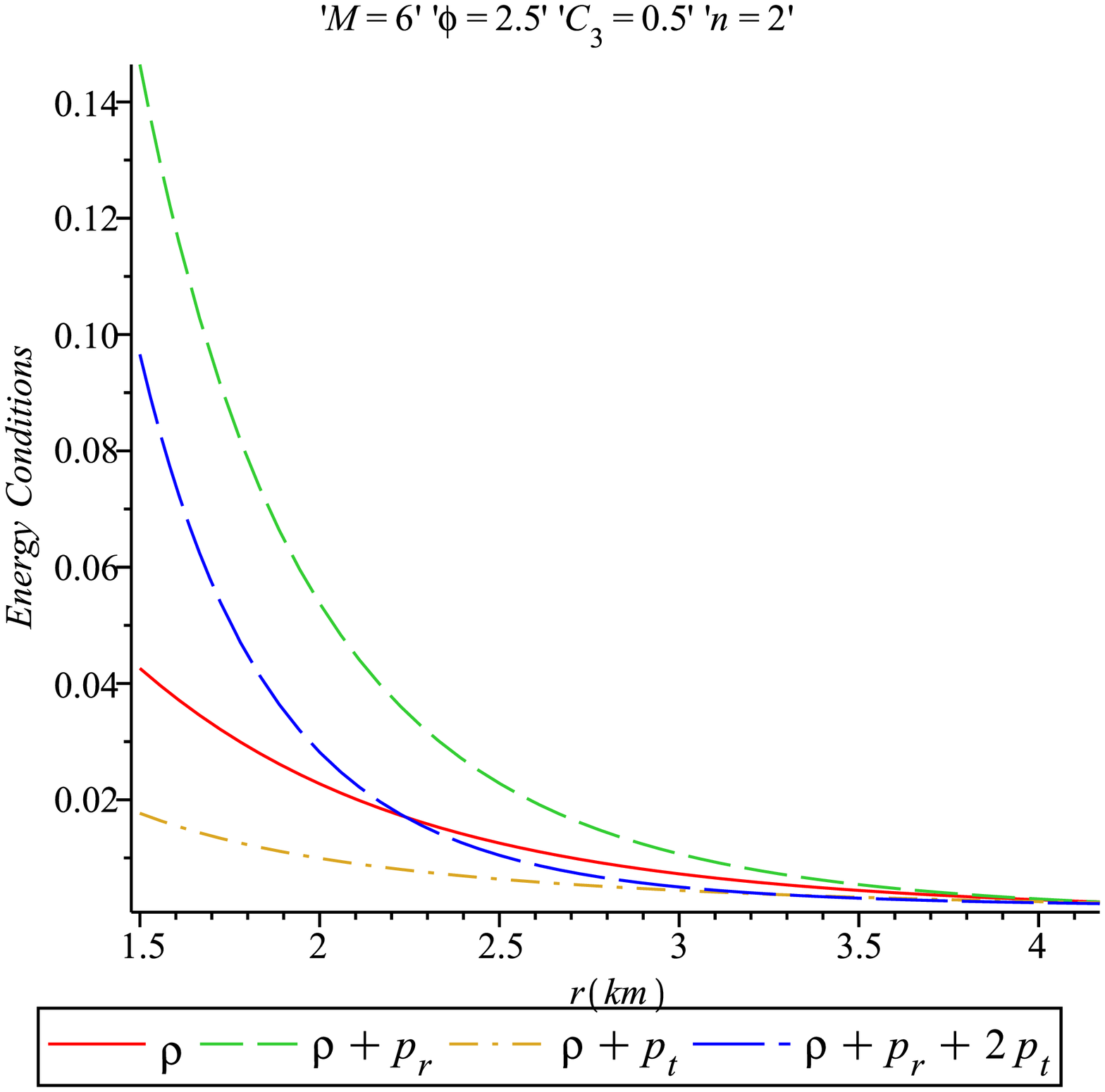}
\includegraphics[width=0.4\textwidth]{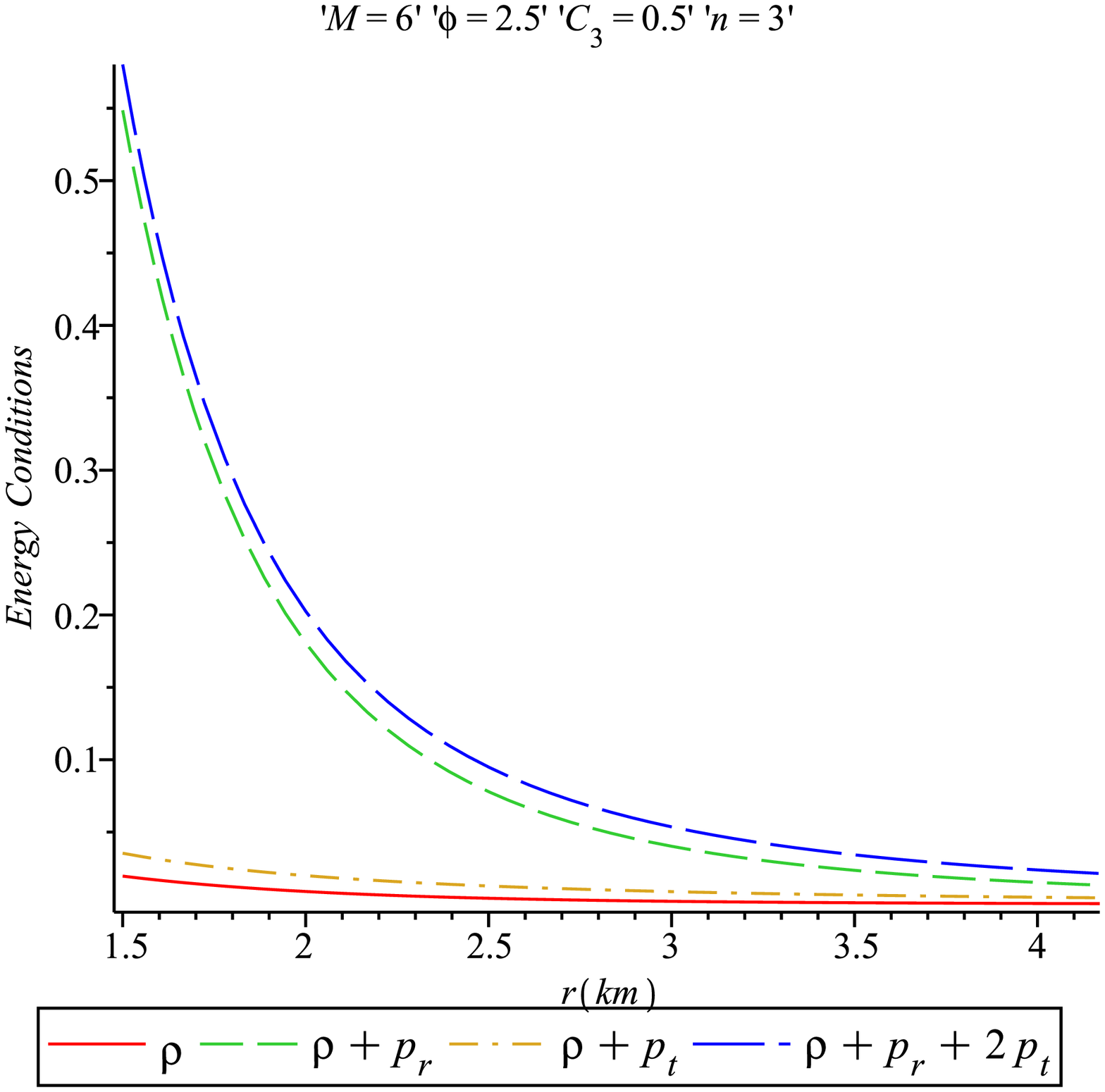}
\includegraphics[width=0.4\textwidth]{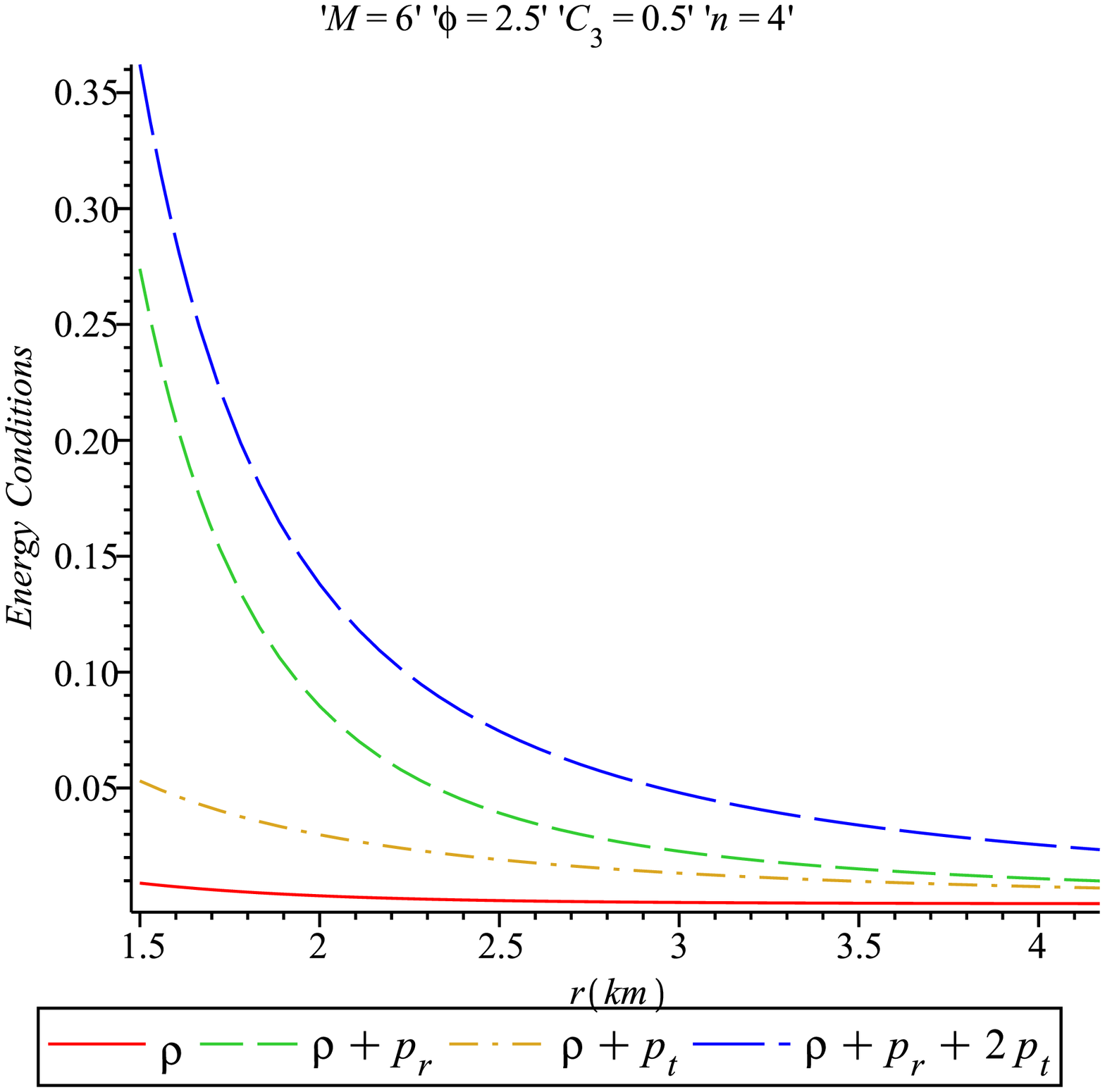}
\includegraphics[width=0.4\textwidth]{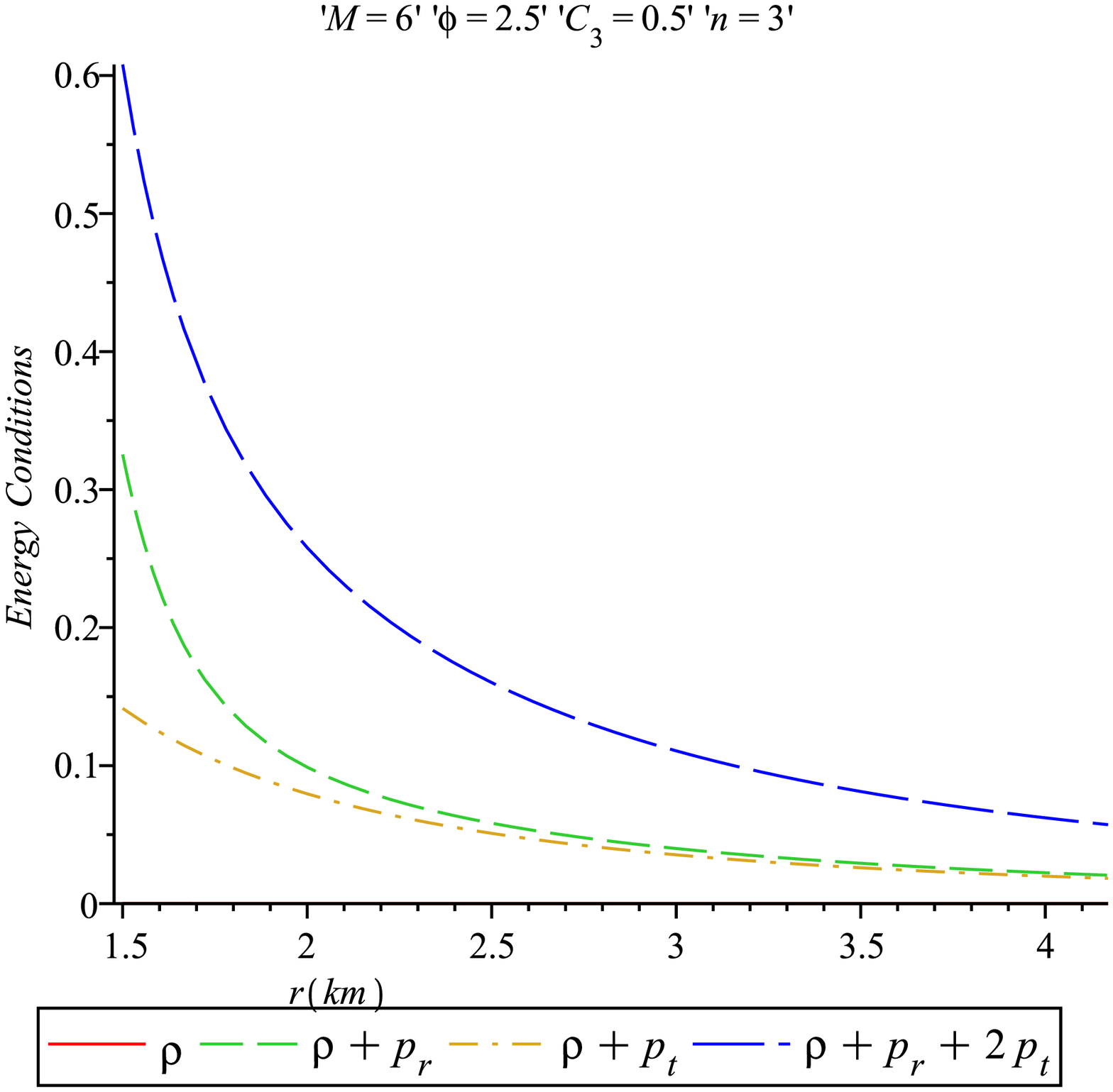}
\caption{The energy conditions in the interior have been
plotted against $r$ for $4D$ spacetime (top left), $5D$ spacetime
(top right), $6D$ spacetime (bottom left) and $11D$ spacetime
(bottom right) }
\end{figure}

\subsection{Anisotropy of the models}
We have shown the possible variation of radial and transverse pressures in Fig. 4
(top left and right of the panel respectively). Hence the measure of anisotropy
$\Delta=(p_t-p_r)$ in $4$, $5$, $6$ and $11$
dimensional cases are respectively given as
\begin{equation}
\Delta_4=\frac{1}{8\pi}\left[\frac{12M\sqrt{\phi}}{\pi
r^{3}}\left\{\frac{1}{\sqrt{\phi}}\arctan\left(\frac{r}{\sqrt{\phi}}\right)-\frac{r}{r^{2}+\phi}\right\}-\frac{1}{r^{2}}
-\frac{8M\sqrt{\phi}}{\pi(r^{2}+\phi)^{2}}-\frac{3A}{C_3^{2}r^{3}}\right],
\end{equation}

\begin{equation}
\Delta_5=\frac{1}{8\pi}\left[\frac{32M\sqrt{\phi}}{3\pi
r^{4}}\left\{\frac{3r^{2}
+2\phi}{(r^{2}+\phi)^{\frac{3}{2}}}\right\}-\frac{1}{r^{2}}
-\frac{8M\sqrt{\phi}}{\pi(r^{2}+\phi)^{\frac{5}{2}}}+\frac{6A}{C_3^{2}r^{4}}\right],
\end{equation}

\begin{equation}
\Delta_6=\frac{1}{8\pi}\left[\frac{5M\sqrt{\phi}}{\pi
r^{5}}\left\{\frac{3}{\sqrt{\phi}}\tan^{-1}\left(\frac{r}{\sqrt{\phi}}\right)
-\frac{5r^{3}+3r\phi}{(r^{2}+\phi)^{2}}
\right\}-\frac{1}{r^{2}}-\frac{8M\sqrt{\phi}}{\pi(r^{2}+\phi)^{3}}-\frac{10A}{r^{5}C_3^{2}}\right],
\end{equation}

\begin{eqnarray}
\Delta_{11}= \frac{1}{8\pi}\left[\frac{M\sqrt{\phi}}{4\pi
r^{10}}\left\{\frac{315r^{8}+840r^{6}\phi+1008r^{4}\phi^{2}+576r^{2}\phi^{3}+128\phi^{4}}{(r^{2}
+\phi)^{\frac{9}{2}}}\right\} \right. \nonumber\\
- \left. \frac{1}{r^{2}}-\frac{8M\sqrt{\phi}}{\pi(r^{2}+\phi)^{\frac{11}{2}}}+\frac{45A}{C_3^{2}r^{10}}\right],
\end{eqnarray}

All these are plotted in Fig. 4 (bottom left of the panel). From all the plots
we see that $\Delta<0$ ~i.e., $p_t~<p_r$ and hence the anisotropic
force is attractive in nature. A detailed study shows that
firstly, in every case of different dimensions the measure of
anisotropy is a decreasing function of $r$. Secondly, from $4D$
onward measure of anisotropy is increasing gradually and
is attaining maximum at $5D$. Surprisingly, it is very high
compared to $4D$ and $11D$ spacetimes. This observation therefore
dictates that $4D$ configuration represents almost a spherical
object as departure from isotropy is very less than the higher
dimensional spacetimes.

Moreover, in all the above cases of different dimension one can
note that the pressure gradient $\frac{dp_r}{dr}$ is a decreasing
function of $r$ (bottom right panel of Fig. 4).

\begin{figure}[h]
\centering
\includegraphics[width=0.4\textwidth]{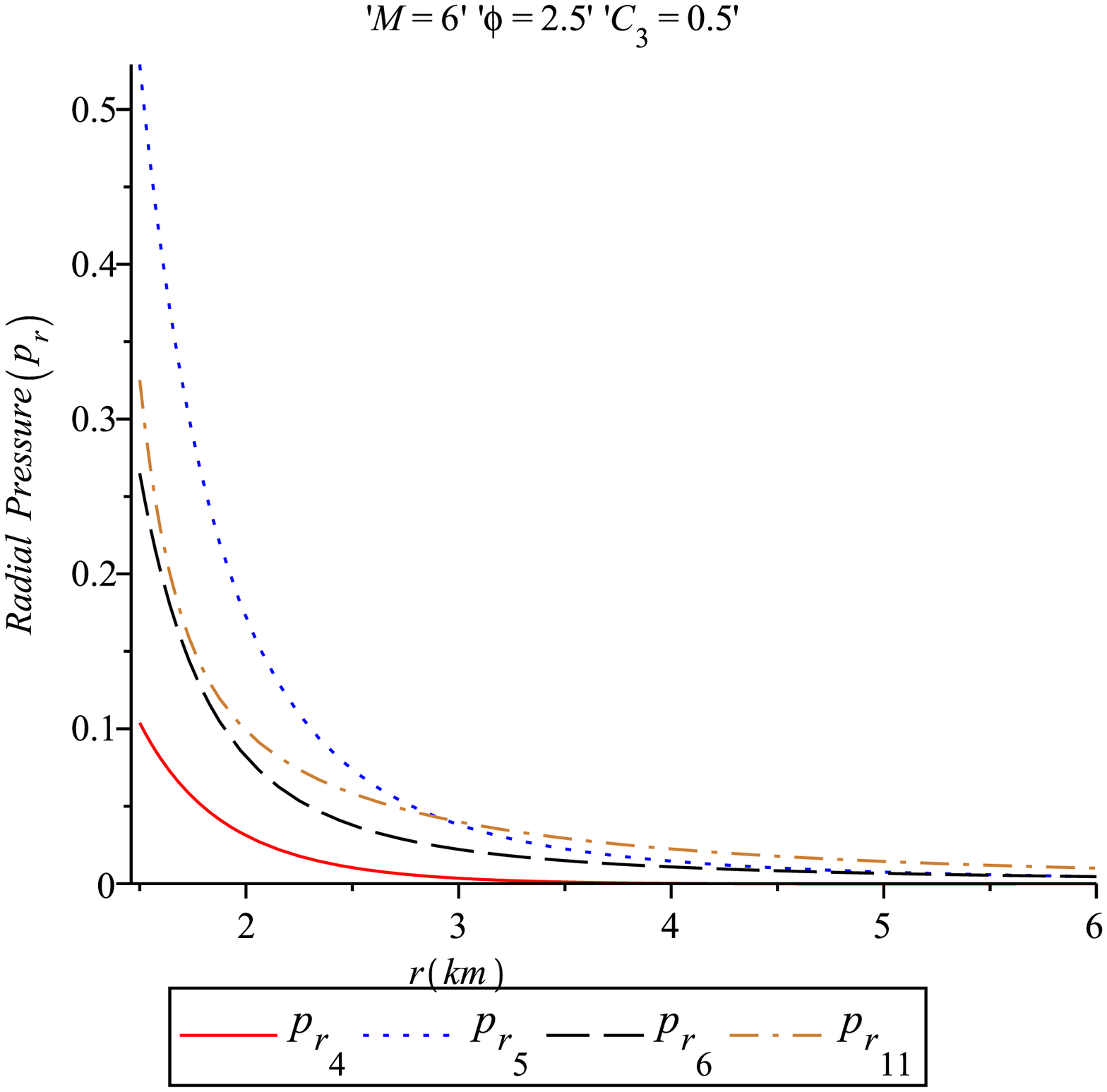}
\includegraphics[width=0.4\textwidth]{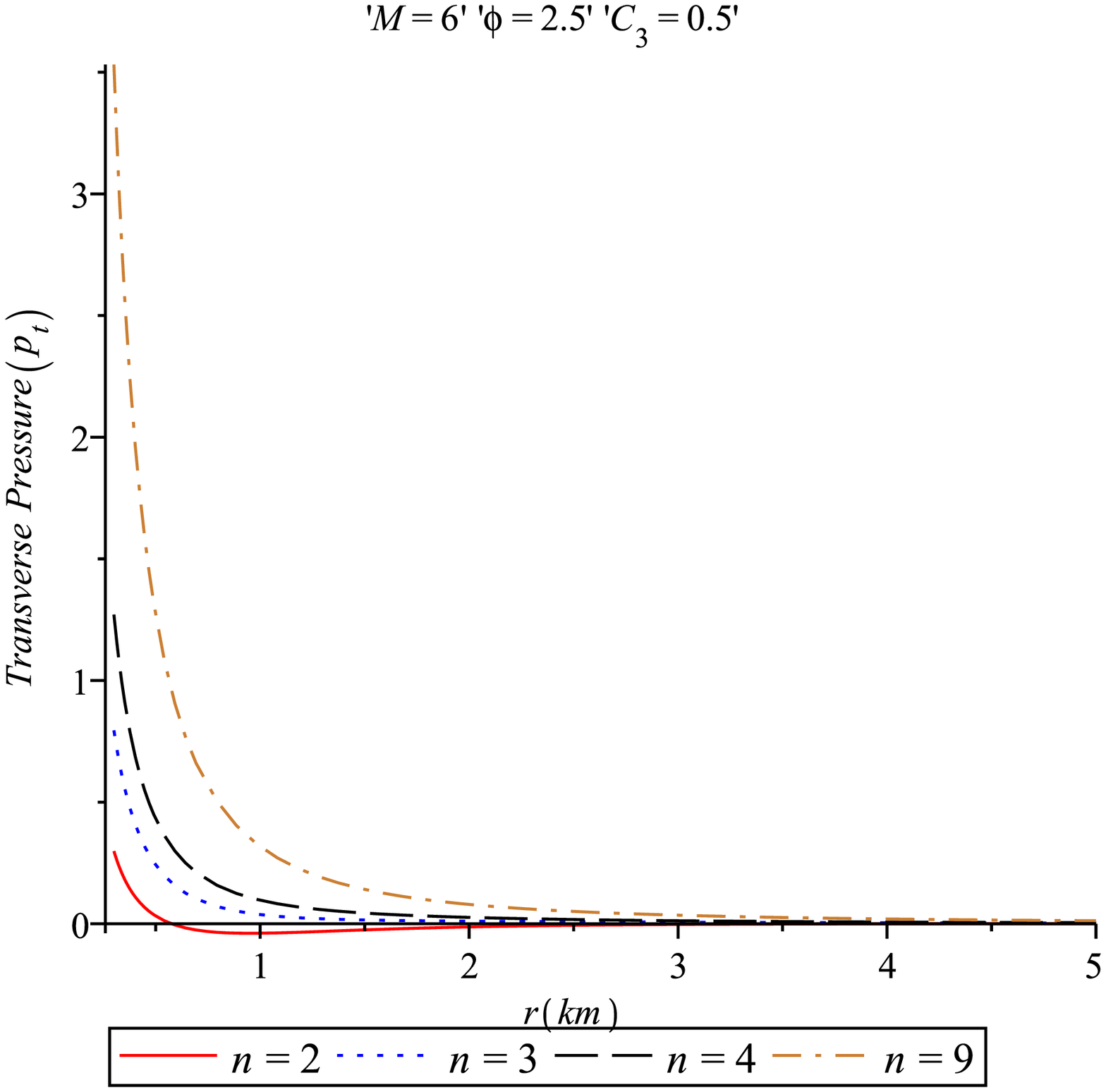}
\includegraphics[width=0.4\textwidth]{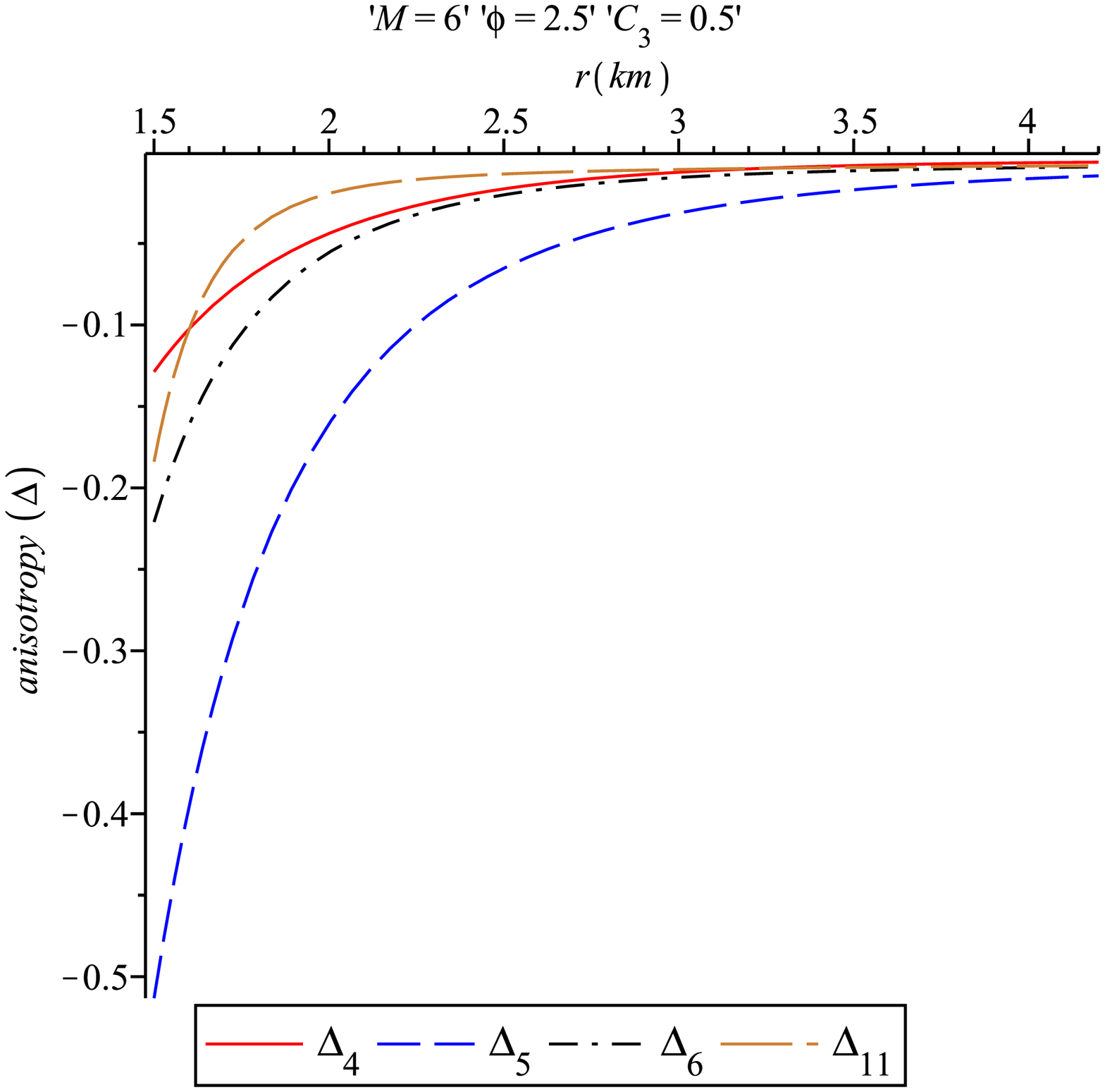}
\includegraphics[width=0.4\textwidth]{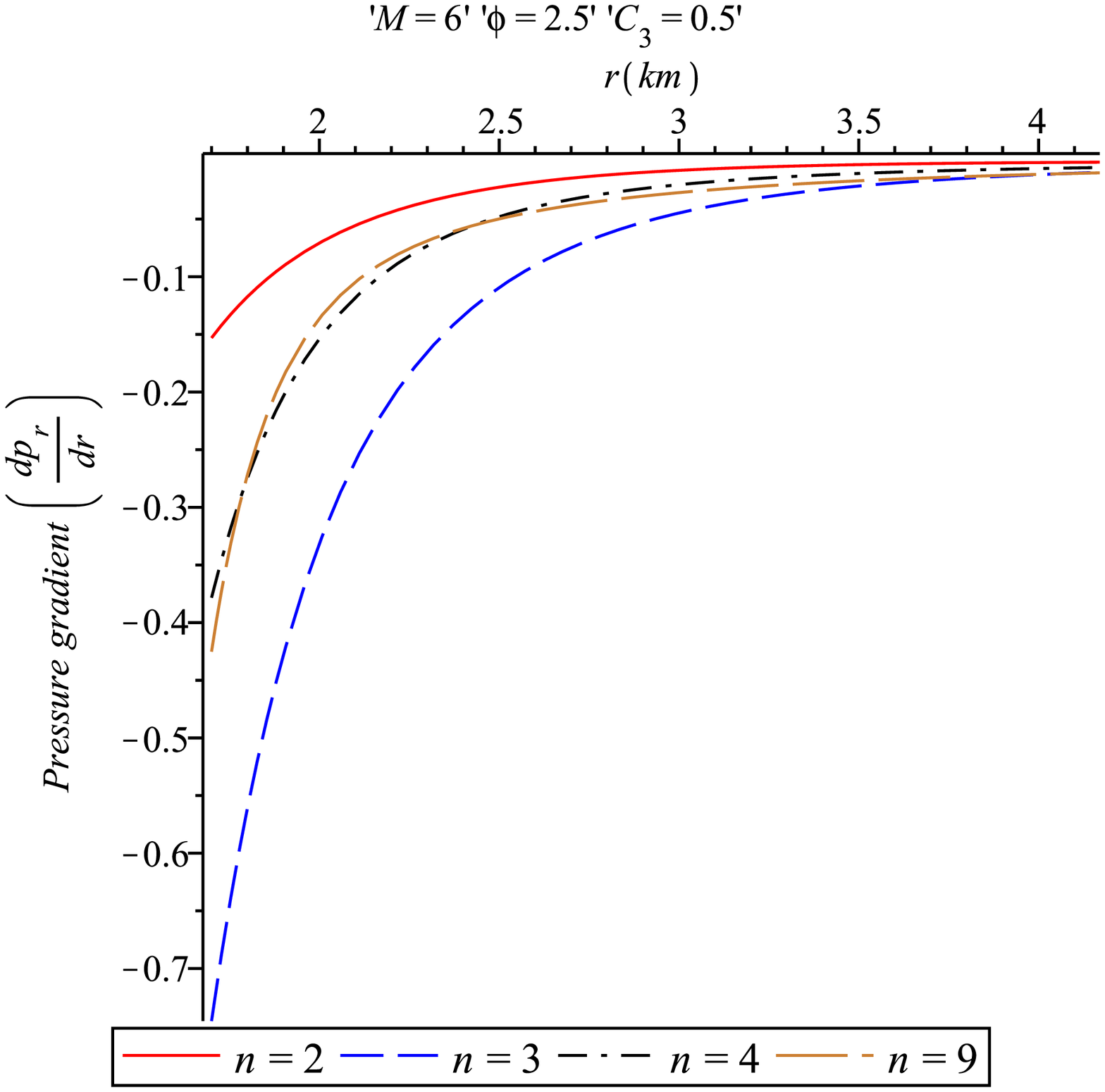}
\caption{Variation of the radial pressure $p_r$ in the interior of
the compact star in $4D$,~$5D$,~$6D$,~$11D$ have been plotted
against $r$ (Left). The figures show that profiles of $p_r$ are
monotonic decreasing function of $r$. The anisotropic factor
$\Delta$ for $4D$,~$5D$,~$6D$,~$11D$ cases are shown against $r$
(right). In both the cases the suffixes in the legends indicate
the dimension of the spacetime}
\end{figure}

\subsection{Compactness and redshift of the star}
At the end of previous Section we did a primary test to get a preliminary
idea about the structure of the star under consideration and we have seen
that the star actually represents a compact object with a radius
$4.17$~km. However, for further test of confirmation one can
perform some specific calculations for `compactness factor'
\cite{Rahaman,Kalam2012,Hossein2012,Kalam2013}.

To do so we first define gravitational mass of the system of
matter distribution as follows:
\begin{equation}
m(r) = \int_0^{ r}~ \left[  \frac{2 \pi^{\frac{n+1}{2}} }{\Gamma
\left(\frac{n+1}{2}\right)}\right]r^n \rho dr.
\end{equation}

Therefore, the compactness factor and surface redshift of the star
can be respectively given by
\begin{equation}
u(r)= \frac{m(r) }{r},
 \end{equation}

\begin{equation}
z_s= [1-2u]^{-1/2}-1.
\end{equation}

Hence for different dimensions we can calculate the expressions for
the above parameters as follows:

\textbf{For n=2:}
\begin{equation}
m(r)=\frac{2M}{\pi}\left[tan^{-1}\left(\frac{r}{\sqrt{\phi}}\right)-\frac{r\sqrt{\phi}}{r^{2}+\phi}\right],
\end{equation}
\begin{equation}
u(r)=\frac{2M}{\pi
r}\left[tan^{-1}\left(\frac{r}{\sqrt{\phi}}\right)-\frac{r\sqrt{\phi}}{r^{2}+\phi}\right],
\end{equation}
\begin{equation}
z_s=\left[1-\frac{4M}{\pi
r}\left\{tan^{-1}\left(\frac{r}{\sqrt{\phi}}\right)-\frac{r\sqrt{\phi}}{r^{2}+\phi}\right\}\right]^{-\frac{1}{2}}-1.
\end{equation}

\textbf{For n=3:}
\begin{equation}
m(r)=\frac{2M}{3}\left[2-\frac{(3r^{2}+2\phi)\sqrt{\phi}}{(r^{2}+\phi)^{\frac{3}{2}}}\right],
\end{equation}
\begin{equation}
u(r)=\frac{2M}{3r}\left[2-\frac{(3r^{2}+2\phi)\sqrt{\phi}}{(r^{2}+\phi)^{\frac{3}{2}}}\right],
\end{equation}
\begin{equation}
z_s=\left[1-\frac{4M}{3r}\left\{2-\frac{(3r^{2}+2\phi)\sqrt{\phi}}{(r^{2}+\phi)^{\frac{3}{2}}}\right\}\right]^{-\frac{1}{2}}-1.
\end{equation}

\textbf{For n=4:}
\begin{equation}
m(r)=M\left[tan^{-1}\left(\frac{r}{\sqrt{\phi}}\right)-\frac{r\sqrt{\phi}(5r^{2}+3\phi)}{3(r^{2}+\phi)^{2}}\right],
\end{equation}
\begin{equation}
u(r)=M\left[\frac{1}{r}tan^{-1}\left(\frac{r}{\sqrt{\phi}}\right)-\frac{\sqrt{\phi}(5r^{2}+3\phi)}{3(r^{2}+\phi)^{2}}\right],
\end{equation}
\begin{equation}
z_s=\left[1-2M\left\{\frac{1}{r}tan^{-1}\left(\frac{r}{\sqrt{\phi}}\right)-\frac{\sqrt{\phi}(5r^{2}+3\phi)}{3(r^{2}+\phi)^{2}}\right\}\right]^{-\frac{1}{2}}-1.
\end{equation}

\textbf{For n=9:}
\begin{equation}
m(r)=\frac{M\pi^{3}}{3780}\left[128-\frac{128\phi^{4}+576\phi^{3}r^{2}+1008\phi^{2}r^{4}
+840\phi
r^{6}+315r^{8}}{(r^{2}+\phi)^{\frac{9}{2}}}\sqrt{\phi}\right],
\end{equation}
\begin{equation}
u(r)=\frac{M\pi^{3}}{3780r}\left[128-\frac{128\phi^{4}+576\phi^{3}r^{2}+1008\phi^{2}r^{4}
+840\phi
r^{6}+315r^{8}}{(r^{2}+\phi)^{\frac{9}{2}}}\sqrt{\phi}\right],
\end{equation}
\begin{equation}
z_s=\left[1-\frac{M\pi^{3}}{1890r}\left\{128-\frac{128\phi^{4}+576\phi^{3}r^{2}+1008\phi^{2}r^{4}
+840\phi
r^{6}+315r^{8}}{(r^{2}+\phi)^{\frac{9}{2}}}\sqrt{\phi}\right\}\right]^{-\frac{1}{2}}-1.
\end{equation}

\begin{figure}[h]
\centering
\includegraphics[width=0.4\textwidth]{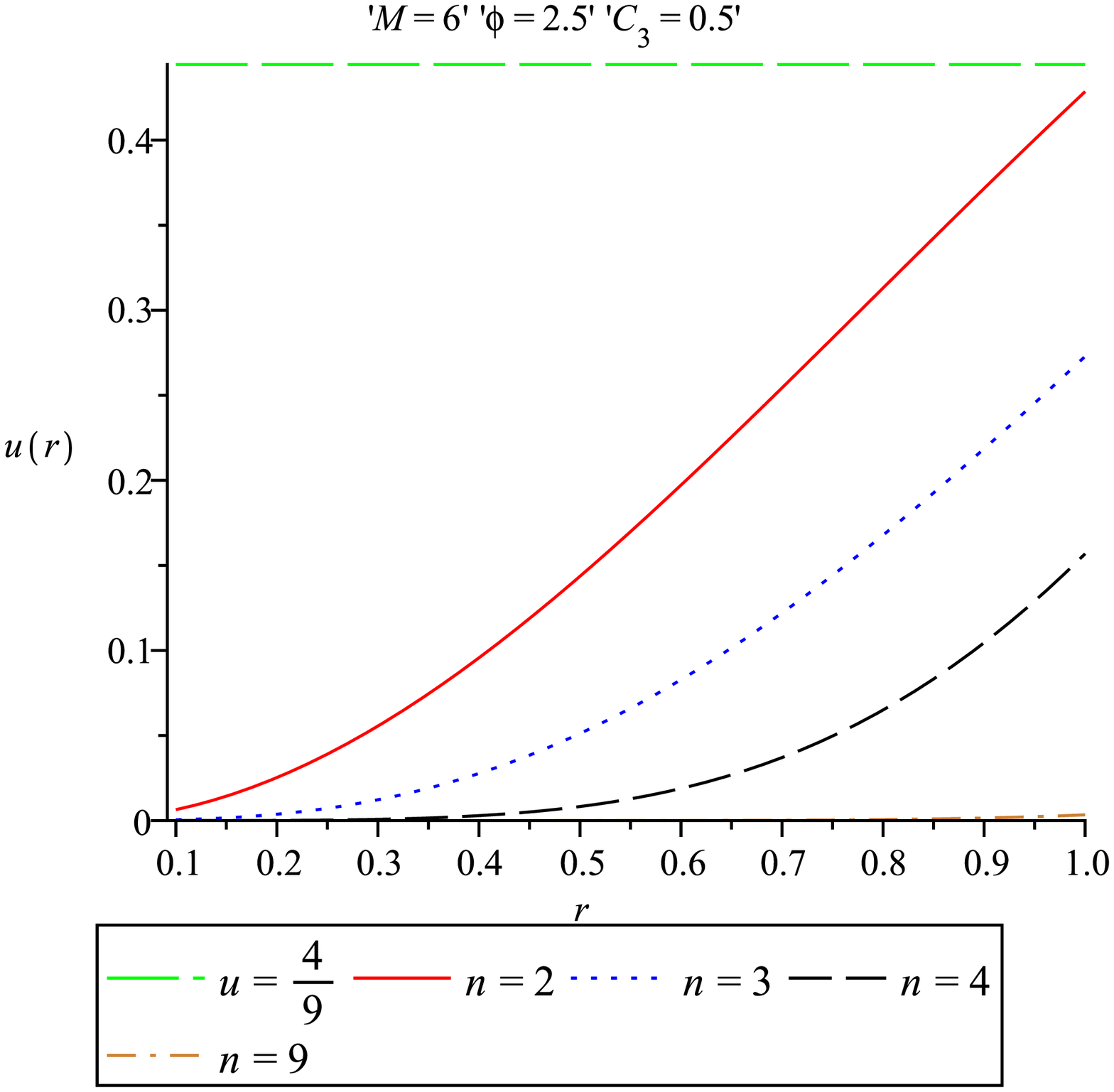}
\includegraphics[width=0.4\textwidth]{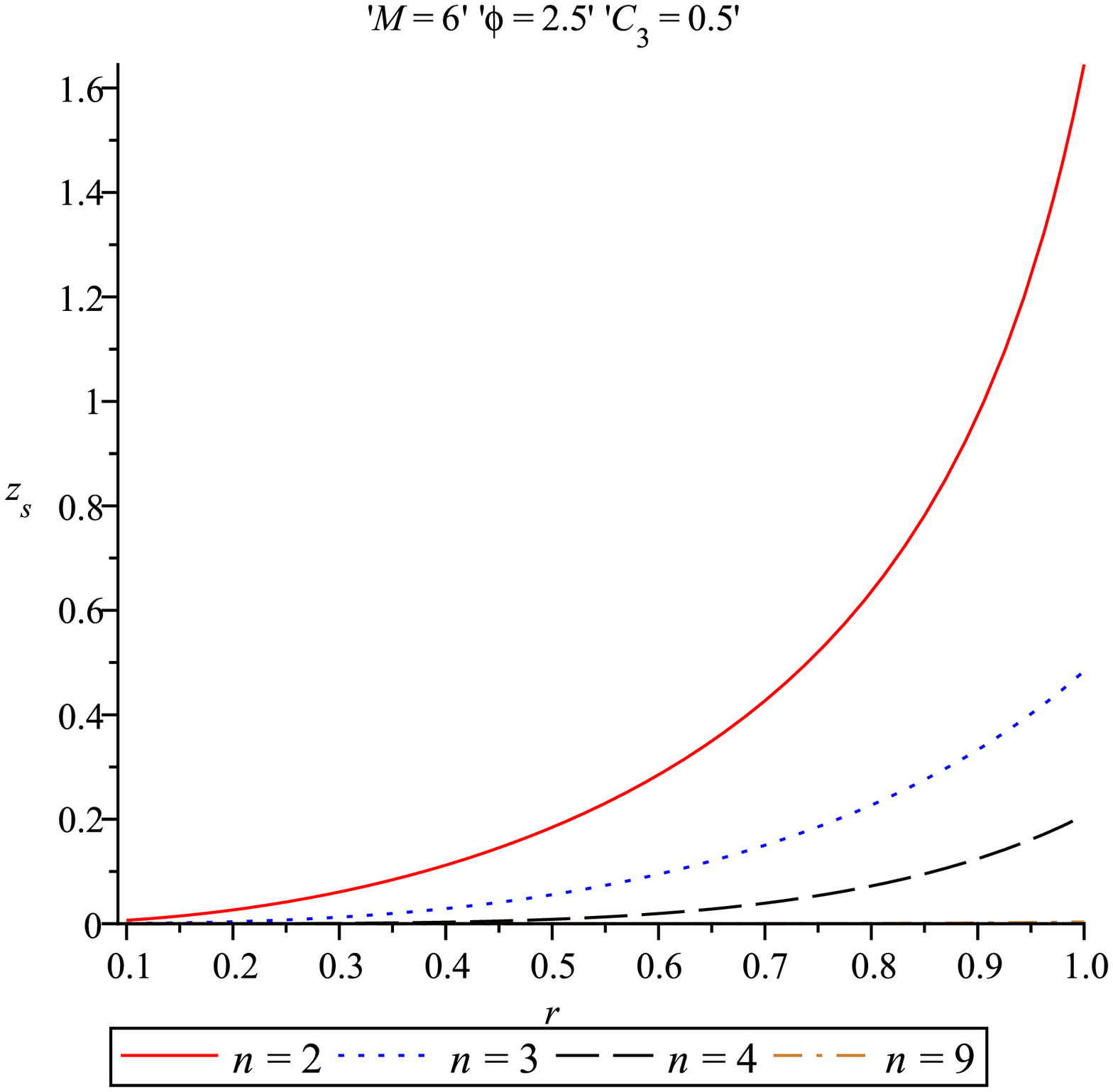}
\caption{The compactness factor (left) and surface redshift (right)
of compact star in $4D$,~$5D$,~$6D$,~$11D$ dimensions are plotted against $r$~(km) for the specified range}
\end{figure}

The nature of variation of the above expressions for compactness
factor and surface redshift of the star can be seen in the Fig. 5
in the left and right panel respectively for all the values of
$n$. It is observed from Fig. 5 (left panel) that compactness
factors for different dimensions are gradually increasing with
decreasing $n$ and maximum for $4D$ spacetime. Thus, very
interestingly, at the center the star is most dense for
$4$-dimension with a very small yet definite core whereas in the
$11D$ case here seem to be no core.

We note that in connection with the isotropic case and in the absence of
the cosmological constant it has been shown for the surface
redshift analysis that $z_s \leq
2$~\cite{Buchdahl1959,Straumann1984,Boehmer2006}. On the other
hand, B{\"o}hmer and Harko \cite{Boehmer2006} argued that for an
anisotropic star in the presence of a cosmological constant the
surface redshift must obey the general restriction $z_s \leq 5$,
which is consistent with the bound $z_s \leq 5.211$ as obtained by
Ivanov~\cite{Ivanov2002}. Therefore, for an anisotropic star
without cosmological constant the above value $z_s \leq 1$ is
quite reasonable as can be seen in the $4D$ case (Fig. 5, right
panel) \cite{Rahaman}. In the other cases of higher
dimension the surface redshift values are increasing and seem to
be within the upper bound~\cite{Ivanov2002}.

We note that integration of $m(r)$ from $0$ to $R$, where $R$
is the radius of the fluid distribution, gives $M$ (total mass of
the source) i.e.
\[M=\int_0^{ R}~ \left[  \frac{2 \pi^{\frac{n+1}{2}} }{\Gamma
\left(\frac{n+1}{2}\right)}\right]r^n \rho dr.\] This equation
gives the radius $R$ of the fluid distribution. Thus solutions of
the following equations provide the corresponding radius of
different dimensional situations.

\textbf{For $n=2$:}
\begin{equation}
M=\frac{2M}{\pi}\left[tan^{-1}\left(\frac{R}{\sqrt{\phi}}\right)
-\frac{R\sqrt{\phi}}{R^{2}+\phi}\right],
\end{equation}

\textbf{For $n=3$:}
\begin{equation}
M=\frac{2M}{3}\left[2-\frac{(3R^{2}+2\phi)\sqrt{\phi}}{(R^{2}+\phi)^{\frac{3}{2}}}\right],
\end{equation}

\textbf{For $n=4$:}
\begin{equation}
M=M\left[tan^{-1}\left(\frac{R}{\sqrt{\phi}}\right)-\frac{R\sqrt{\phi}(5R^{2}+3\phi)}
{3(R^{2}+\phi)^{2}}\right],
\end{equation}

\textbf{For $n=9$:}
\begin{equation}
M=\frac{M\pi^{3}}{3780}\left[128-\frac{128\phi^{4}+576\phi^{3}R^{2}+1008\phi^{2}R^{4}
+840\phi
R^{6}+315R^{8}}{(R^{2}+\phi)^{\frac{9}{2}}}\sqrt{\phi}\right],
\end{equation}

\subsection{Some other physical parameters}
In this subsection we have shown the panel of the plots for the
conformal parameter $\psi(r)$ (top left), the metric potential
$e^{\lambda}$ (top right) and the density $\rho$ (bottom) for $4$
and extra dimensional spacetimes (Fig. 6). It is observed that for
all the physical parameters the features are as usual for $4D$,
however for extra dimension they take different shapes. A special
mention can be done for density where central densities are
abruptly decreasing as one goes to higher dimensions. Thus, from
the plot it reveals that the central density is maximum for $4D$
whereas it is minimum for $11D$ spacetime showing most compactness
of the star for standard $4$-dimension. Note that this same result
was observed in Fig 5 (left panel).

\begin{figure}[h]
\centering
\includegraphics[width=0.4\textwidth]{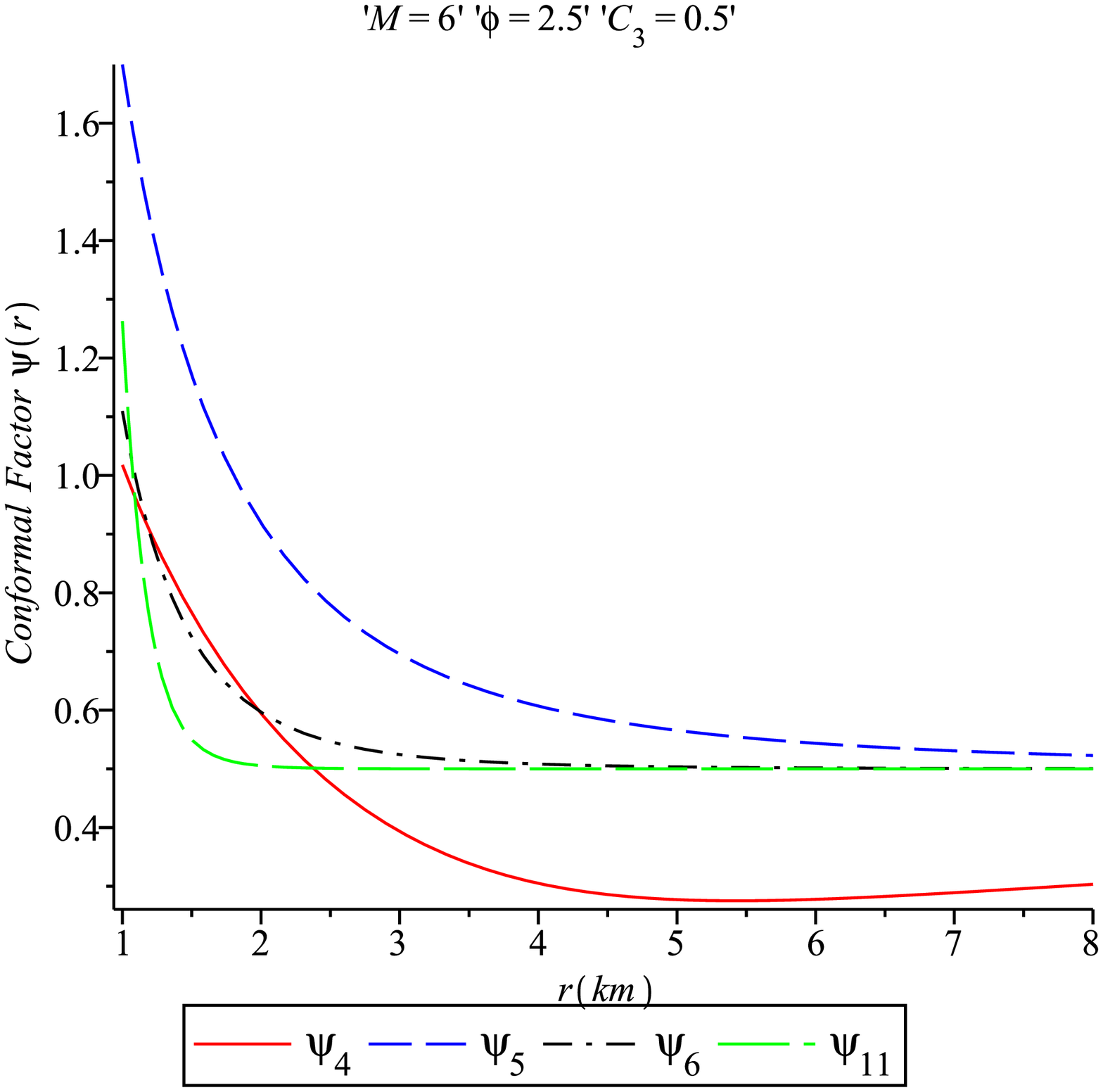}
\includegraphics[width=0.4\textwidth]{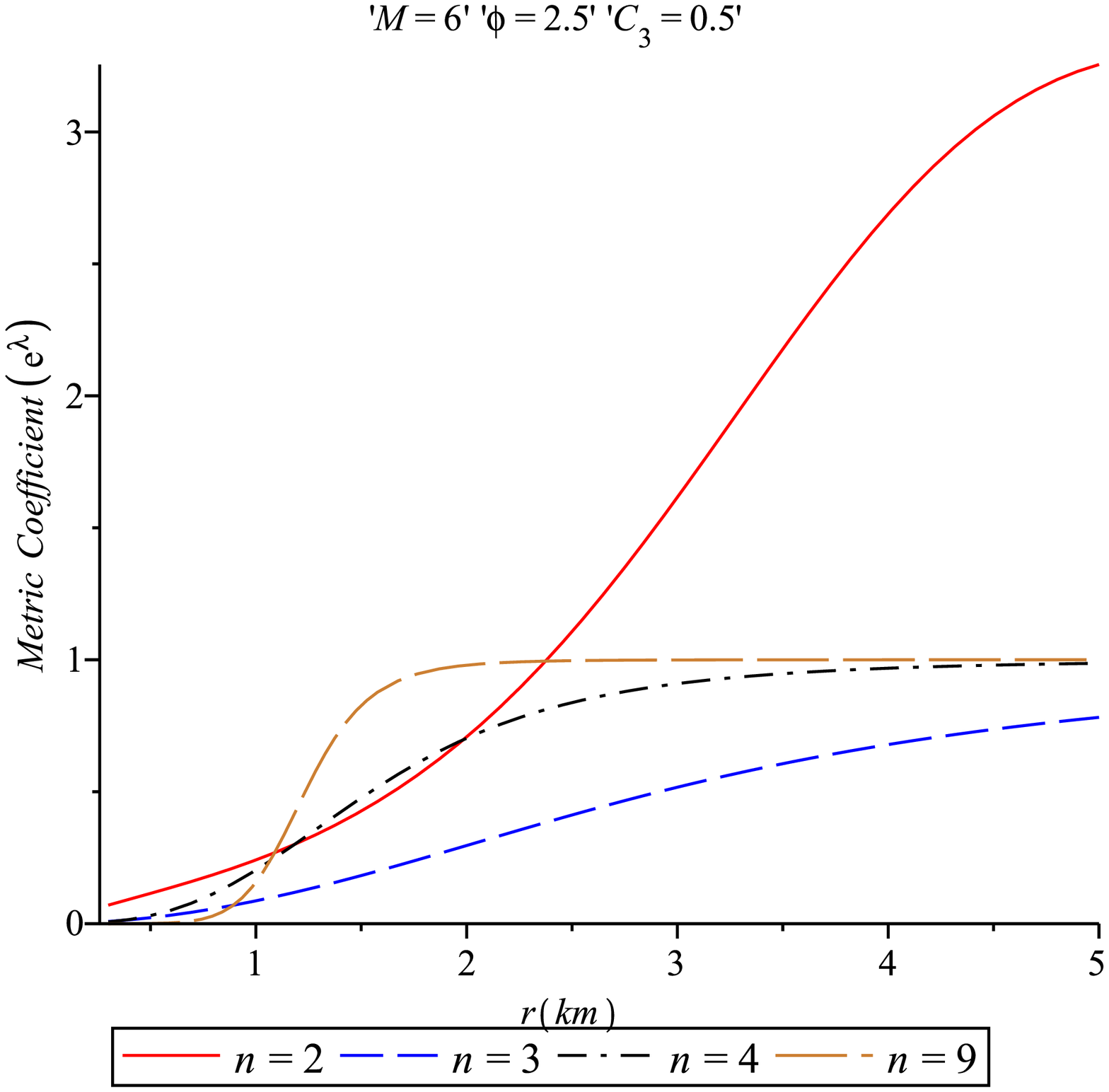}
\includegraphics[width=0.4\textwidth]{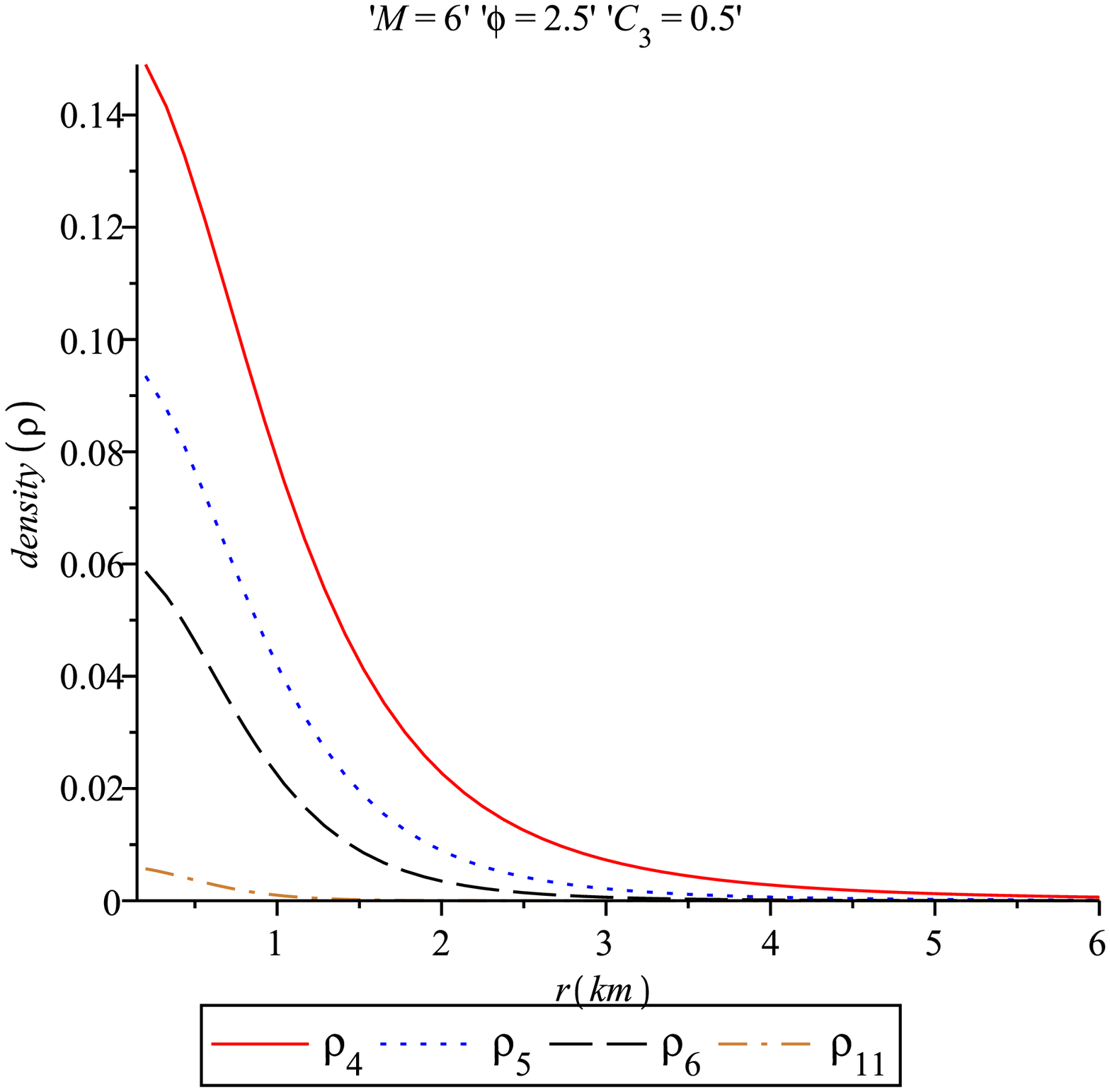}
\caption{The conformal factor $\psi(r)$ (top left),
the metric potential $e^{\lambda}$ (top right) and matter density (bottom)
are plotted against $r$~(km) for $4D$,~$5D$,~$6D$,~$11D$ dimensional spacetimes.
The suffixes in the left panel indicates the dimension of the spacetime
and for right panel $D=n+2$ stands for the dimension of the spacetime}
\end{figure}

\section{Conclusion}
In the present paper we have studied thoroughly a set of new
interior solutions for anisotropic stars admitting conformal motion
in higher dimensional noncommutative spacetime. Under this
spacetime geometry the Einstein field equations are solved by
choosing a particular Lorentzian type density distribution
function as proposed by Nozari and Mehdipour \cite{Nozari2009}. The
studies are conducted not only with standard $4D$ dimensional
spacetime but also for three special cases with higher dimension,
such as $5D$, $6D$ and $11D$. In general it is noted that the
model parameters e.g. matter-energy density, radial as well as
transverse pressures, anisotropy and others show physical
behaviours which are mostly regular throughout the stellar
configuration.

Also it is specially observed that the solutions represent a star
with mass $2.27$~$M_{\odot}$ and radius $4.17$~km
which falls within the range ($0 < z_s \leq 1$) of a compact
star \cite{Rahaman,Kalam2012,Hossein2012,Kalam2013}.
However, it has been shown that for a strange star of radius
$6.88$~km surface redshift turns out to be $z_s = 0.5303334$
\cite{Rahaman} whereas the maximum surface redshift for a
strange star $Her X-1$ of radius $7.7$~km is $0.022$
\cite{Kalam2012} and that for a compact star $4U~1820-30$ of
radius $10$~km turns out to be again $0.022$
\cite{Hossein2012}. Therefore it seems that our compact star
may be a strange quark star (see Table 1).

\begin{table*}
\centering
\begin{minipage}{140mm}
\caption{Values of the model parameter $z_s$ for
different Strange Stars}\label{tbl-1}
\begin{tabular}{@{}lrrrrrr@{}}
\hline Strange Star candidates    & $M$($M_{\odot}$) & $R$(km) &
$M/R$         & $z_s$ \\ \hline Her X-1                   & 0.88 &
7.7     & 0.168         & 0.0220 [Ref. 13]\\
                          &                  &         &               & 0.2285 [Ref. 15]\\
4U 1820-30                & 2.25             & 10.0    & 0.332 &
0.0220 [Ref. 14]\\
                          &                  &         &               & 0.7246 [Ref. 15]\\
SAX J 1808.4-3658(SS1)    & 1.435            & 7.07    & 0.299 &
0.5787 [Ref. 15]\\ SAX J 1808.4-3658(SS2)    & 1.323 & 6.35    &
0.308         & 0.6108 [Ref. 15]\\ Rahaman model [Ref. 11]    &
1.46             & 6.88    & 0.313         & 0.5303334
\\ Our proposed model        & 2.27             & 4.17    & 0.804
& $0 < z_s \leq 1$\\ \hline
\end{tabular}
\end{minipage}
\end{table*}

However, through several mathematical case studies we have given
emphasis on the acceptability of the model from physical point of
view for various structural aspects. As a consequence it is
observed that for higher dimensions, i.e. beyond $4D$ spacetime,
the solutions exhibit several interesting yet bizarre features.
These features seem physically not very unrealistic.

Thus, as a primary stage, the investigation indicates that compact
stars may exist even in higher dimensions. But before placing a
demand in favour of this highly intrigued issue of compact stars
with extra dimensions we need to perform more specific studies and
to look at the diversified technical aspects related to higher
dimensional spacetimes of a compact star. Basically our approach,
dependent on a particular energy density distribution of
Lorenztian type, which gives higher dimensional existence of
compact stars may not be only the way to have sufficient evidence
in favour of it. We further need to employ other type of density
distributions as well. Moreover, one may also think for other than
higher dimensional embedding of GTR and thus opt for alternative
theories of gravity to find conclusive proof for higher
dimensional compact stars.

However, in the literature there are some evidences available
in favour of `Extra Dimensions in Compact Stars'
\cite{Liddle1990,BLL2003,Paul2004,BLL2006,BLL2010,CP2010}.

\section*{Acknowledgments}
FR and SR wish to thank the authorities of the Inter-University
Centre for Astronomy and Astrophysics (IUCAA), Pune, India for
providing the Visiting Associateship under which a part of this
work was carried out. We all express our grateful thanks to both
the referees for their several suggestions which have enabled us
to improve the manuscript substantially.

\end{document}